\documentclass[pre,twocolumn,superscriptaddress]{revtex4}  

\usepackage[paperwidth=210mm,paperheight=297mm,centering,hmargin=2cm,vmargin=2.5cm]{geometry}

\usepackage[pdftex]{graphicx}
\usepackage{amsmath,amsfonts,amssymb}
\usepackage{morefloats}

\usepackage{hyperref}
\usepackage{xcolor} 

\definecolor{bluemoi}{rgb}{0.25,0.50 ,0.75} 

\hypersetup{
  bookmarksopen=false,
  pdftitle=" ",
  pdfauthor=" ", 
  pdfsubject=" ", 
  pdftoolbar=true, 
  pdfmenubar=true, 
  pdfhighlight=/O, 
  colorlinks=true, 
  pdfpagemode=UseNone, 
  pdfpagelayout=SinglePage, 
  pdffitwindow=true, 
  linkcolor=bluemoi, 
  citecolor=bluemoi, 
  urlcolor=bluemoi 
}

\makeatletter
\renewcommand{\figurename}{Figure}
\makeatother
\makeatletter
\renewcommand{\fnum@figure}{\small\textbf{\figurename~\thefigure}}
\makeatother

\begin{document}

\title{Influence of sociodemographic characteristics on human mobility}

\author{Maxime Lenormand}\affiliation{Instituto de F\'isica Interdisciplinar y Sistemas Complejos IFISC (CSIC-UIB), Campus UIB, 07122 Palma de Mallorca, Spain}

\author{Thomas Louail}\affiliation{Institut de Physique Th\'{e}orique, CEA-CNRS (URA 2306), F-91191, 
Gif-sur-Yvette, France}\affiliation{G\'eographie-Cit\'es, CNRS-Paris 1-Paris 7 (UMR 8504), 13 rue du
  four, FR-75006 Paris, France}
	
\author{Oliva G. Cant\'u-Ros}\affiliation{Nommon Solutions and Technologies, calle Ca\~nas 8, 28043 Madrid, Spain}

\author{Miguel Picornell}\affiliation{Nommon Solutions and Technologies, calle Ca\~nas 8, 28043 Madrid, Spain}

\author{Ricardo Herranz}\affiliation{Nommon Solutions and Technologies, calle Ca\~nas 8, 28043 Madrid, Spain}

\author{Juan Murillo Arias}\affiliation{BBVA Data \& Analytics, Avenida de Burgos 16D, 28036 Madrid, Spain}

\author{Marc Barthelemy}\affiliation{Institut de Physique Th\'{e}orique, CEA-CNRS (URA 2306), F-91191, 
Gif-sur-Yvette, France}\affiliation{Centre d'Analyse et de Math\'ematique Sociales, EHESS-CNRS (UMR
8557), 190-198 avenue de France, FR-75013 Paris, France}

\author{Maxi San Miguel}\affiliation{Instituto de F\'isica Interdisciplinar y Sistemas Complejos IFISC (CSIC-UIB), Campus UIB, 07122 Palma de Mallorca, Spain}

\author{Jos\'e J. Ramasco}\affiliation{Instituto de F\'isica Interdisciplinar y Sistemas Complejos IFISC (CSIC-UIB), Campus UIB, 07122 Palma de Mallorca, Spain}

\begin{abstract} 
Human mobility has been traditionally studied using surveys that deliver snapshots of population displacement patterns. The growing accessibility to ICT information from portable digital media has recently opened the possibility of exploring human behavior at high spatio-temporal resolutions. Mobile phone records, geolocated tweets, check-ins from Foursquare or geotagged photos, have contributed to this purpose at different scales, from cities to countries, in different world areas. Many previous works lacked, however, details on the individuals' attributes such as age or gender. In this work, we  analyze credit-card records from Barcelona and Madrid and by examining the geolocated credit-card transactions of individuals living in the two provinces, we find that the mobility patterns vary according to gender, age and occupation. Differences in distance traveled and travel purpose are observed between younger and older people, but,  curiously, either between males and females of similar age. While mobility displays some generic features, here we show that sociodemographic characteristics play a relevant role and must be taken into account for mobility and epidemiological modelization.
\end{abstract}

\maketitle

\section{INTRODUCTION}

Everyday, billions of individuals generate a large volume of geolocated data by using their mobile phone, GPS, public transport cards or credit cards. Such a vast amount of data is bringing new opportunities for the research in socio-technical systems \cite{Watts2007,Lazer2009,Vespignani2009}. Indeed, geolocated data allow the identification of when and where people interact with or through ICT tools. Each time someone makes a  phone call or pays with a credit card the event gets registered contributing to massive databases with potential to provide useful insights on human behavior and mobility \cite{chowell03,barrat04,Brockmann2006,Gonzalez2008,Song2010,Bagrow2012}. For example,  the authors of Refs. \cite{Brockmann2006,Gonzalez2008} used credit card and mobile phone datasets to study statistical characteristics of mobility patterns and showed that the distribution of displacement of all users can be approximated by a Levy law. Recently, geolocated data has been also employed to study the spatial structure of cities by detecting hotspots \cite{Louail2014} or to characterize land use patterns in urban areas \cite{Ratti2006, Reades2007,Soto2011,Pei2013,Toole2014} with mobile phone records, Twitter data \cite{Frias2012} or both together \cite{Lenormand2014a}. On a larger scale, comparisons and relations between different cities \cite{Noulas2012} or even between countries \cite{Hawelka2013,Lenormand2014b} have also been also investigated.

Beyond mere location, some datasets offer the opportunity to gather extra information about the type and duration of the interaction or the operation through ICT tools. For instance, it is possible to know from mobile phone records where and when an individual makes a call, but sometimes information such as the ID of the callee and the call duration are also available. This information enables researchers to move further on the study of human behavior by analyzing the structure, intensity and spatial properties of social interactions. Some examples include the analysis of the structure of social networks \cite{Liben2005,Onnela2007,Java2007,Huberman2008,Eagle2009,Ferrara2012,Grabowicz2013}, the correlation between mobility and social network \cite{Backstrom2010,Calabrese2011,Phithakkitnukoon2012}, information diffusion \cite{Ferrera2013} and the role played by social groups \cite{Grabowicz2012,Ferrara2012}.

\begin{figure*}
\centering
\includegraphics[scale=0.4]{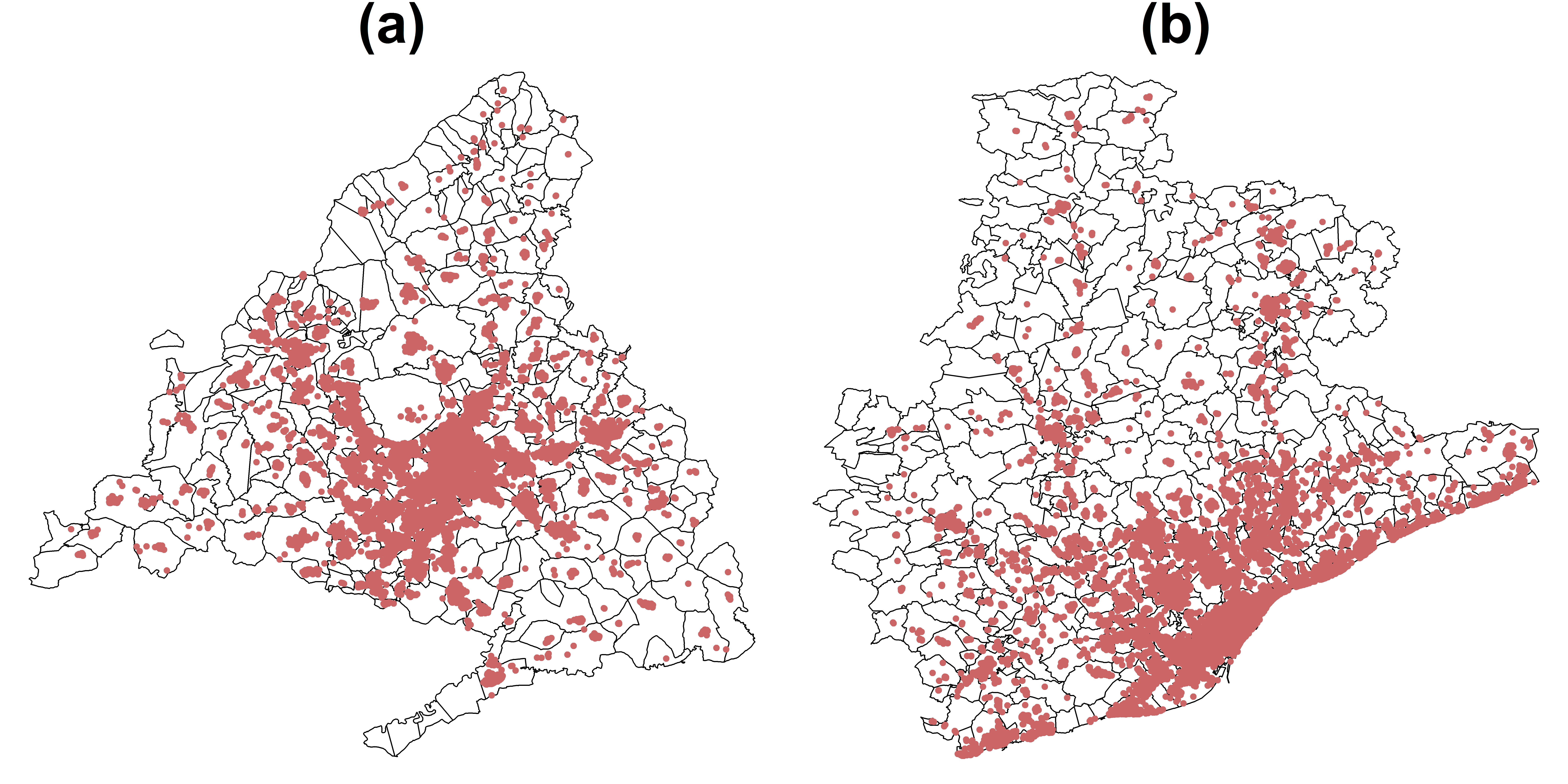}
\caption{\textbf{Maps of the transactions.} The red dots represent the locations of the transactions on a map of the province of Madrid (a) and Barcelona (b). The small areas correspond to postcodes. \label{Fig1}}
\end{figure*}

However, many previous studies lack sociodemographic resolution on the characteristics of the individuals. Except for some features such as language or place of work and/or residence identified in \cite{Hawelka2013,Mocanu2013}, information about gender, age or occupation are typically missing from studies based on ICT data. This information is of great relevance to characerize the city structure, to estimate population needs in urban planning, transport demand and also for public health. For example, regarding age, knowing the areas of concentration of younger and older population helps to optimize infrastructure such as location of schools, care facilities, etc.  
Another aspect for which this information is relevant is the modeling of infectious diseases spreading. The models rely on the interplay among hosts, which is related to their location and mobility. Recent epidemic modeling has incorporated mobility information as a way to get closer to real disease spreading \cite{chowell03,rvachev85,grais03,eubank04,hufnagel04,longini05,ferguson05,riley07,colizza07,ajelli08, balcan09,bajardi11,meloni11,tizzoni12,poletto12}. Additionally, demographic factors such as age or gender can also play an important role in disease transmission and, therefore, must be taken into account when modeling certain infections \cite{wallinga06,nishiura10,nishiura11,rocha11,apolloni13,apolloni14}. Furthermore, in a sort of feedback loop, these sociodemographic factors influence mobility as well. 

Some works based on smaller-scale surveys point out towards a number of significant differences between men and women in terms of their travel purposes and the activities they pursue \cite{Golob1997,Hamed1993,Bianco1996}. More recently, quantitative studies of social networks dynamics have also shown that people behave differently according to the gender and age \cite{McPherson2001,Stehle2013}.  In this paper, we go beyond by analyzing a credit card use database containing over $40$ million card transactions in order to explore consumption and mobility patterns of bank customers in the two most populated provinces of Spain according to three sociodemographic characteristics: gender, age and occupation. 

\vspace*{-0.5cm}
\section{MATERIALS AND METHODS}

\subsection{Dataset description}

Our dataset comes from an extraction of the Banco Bilbao Vizcaya Argentaria (BBVA) database on credit card transactions. Different extractions of this data have been used in open data challenges \cite{BBVA2013} and other scientific works \cite{Sobolevsky2014}. The data contains information about $40$ million bank card transactions made in the provinces of Madrid and Barcelona in $2011$. Each transaction is characterized by its amount (in euro currency) and the time when the transaction has occurred. Each transaction is also linked to a customer and a business using anonymized customer and business IDs. Customers are identified with an anonymized customer ID connected with sociodemographic characteristics (gender, age and occupation) and the postcode of his/her place of residence. For convenience sake, we consider five age groups ($]15,30]$, $]30,45]$, $]45,60]$, $]60,75]$, $>75$) and five types of occupations (student, unemployed, employed, homemaker, and retired). In the same way, businesses are identified with an anonymized business ID, a business category (accommodation, automotive industry, bars and restaurants, etc.) and the geographical coordinates of the credit card terminal. 

The geographical extent of our data is restricted to the provinces of Barcelona and Madrid. For both case studies, we only consider the credit card payments made in the province by individuals living in the province (Figure \ref{Fig1}). Table \ref{tab1} presents some basic statistics on the data collected. Both provinces have similar features in terms of population size, area and number of businesses, but the number of users and transactions are higher in Madrid than in Barcelona. The number of users represents about $8\%$ of the total census population in Madrid and $5\%$ of that of Barcelona.

\begin{table}[!ht]
	\caption{Summary statistics of the two provinces}
	\label{tab1}
		\begin{center}
			\begin{tabular}{lcc}
				\hline
				\centering Statistics   & Barcelona &  Madrid      \\
				\hline
				Number of postcodes  &   368  &  271            \\
				Number of inhabitants  & 5,540,925 &  6,489,680      \\
				Area (km$^2$)         & 7,733    &  8,022            \\
				Number of customers    & 270,205  &  531,818         \\
				Number of transactions   & 13,077,178  &  24,920,896   \\
				Number of businesses   &  111,956  &  109,707         \\ 
				\hline
	  	\end{tabular}
	  \end{center}
\end{table}

\begin{figure*}
\centering
\includegraphics[scale=0.7]{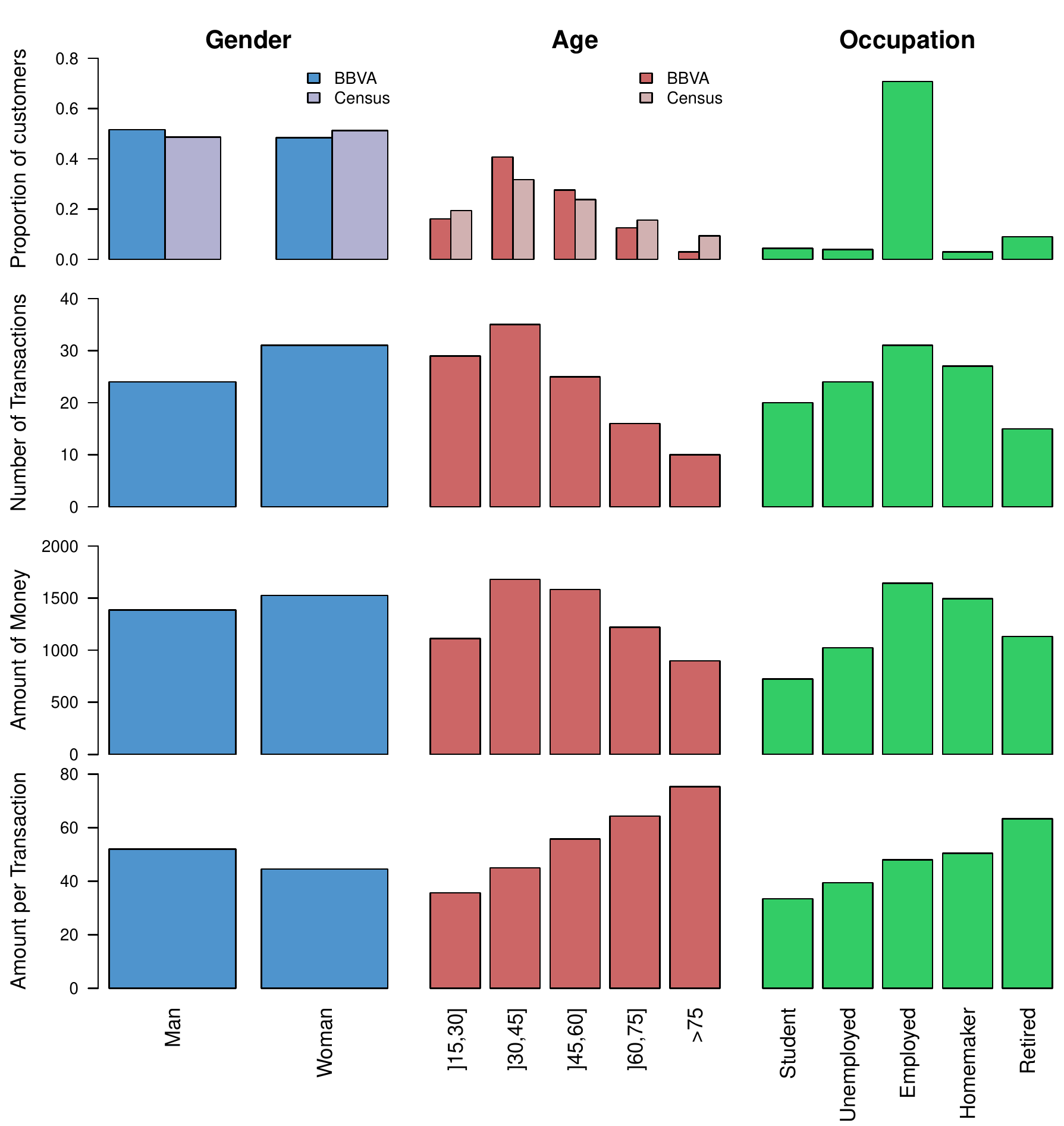}
\caption{\textbf{Descriptive statistics according to the individual sociodemographic characteristics.} From top to bottom, proportion of individuals, median number of transactions per user and per year, median amount of money spent per user and per year (in euro) and median of the average amount of money spent per transaction (in euro) according to, from left to right, the gender, the age and the occupation. \label{Fig2}}
\end{figure*} 

\section{RESULTS}

The statistical features of the data for Barcelona and Madrid are very similar.  Therefore, the data is aggregated for analyzing general properties in the next two sections and segregated later in the third one to study mobility patterns. The aggregation provides higher statistical power, while the disaggregation is needed due to the different geographical shapes of both provinces. Due to the optimization of space,  only figures obtained for Madrid are displayed in the third section on mobility. Still equivalent results for Barcelona are found and can be seen in appendix (Figures S9 - S15).

\begin{figure*}
\centering
\includegraphics[scale=0.7]{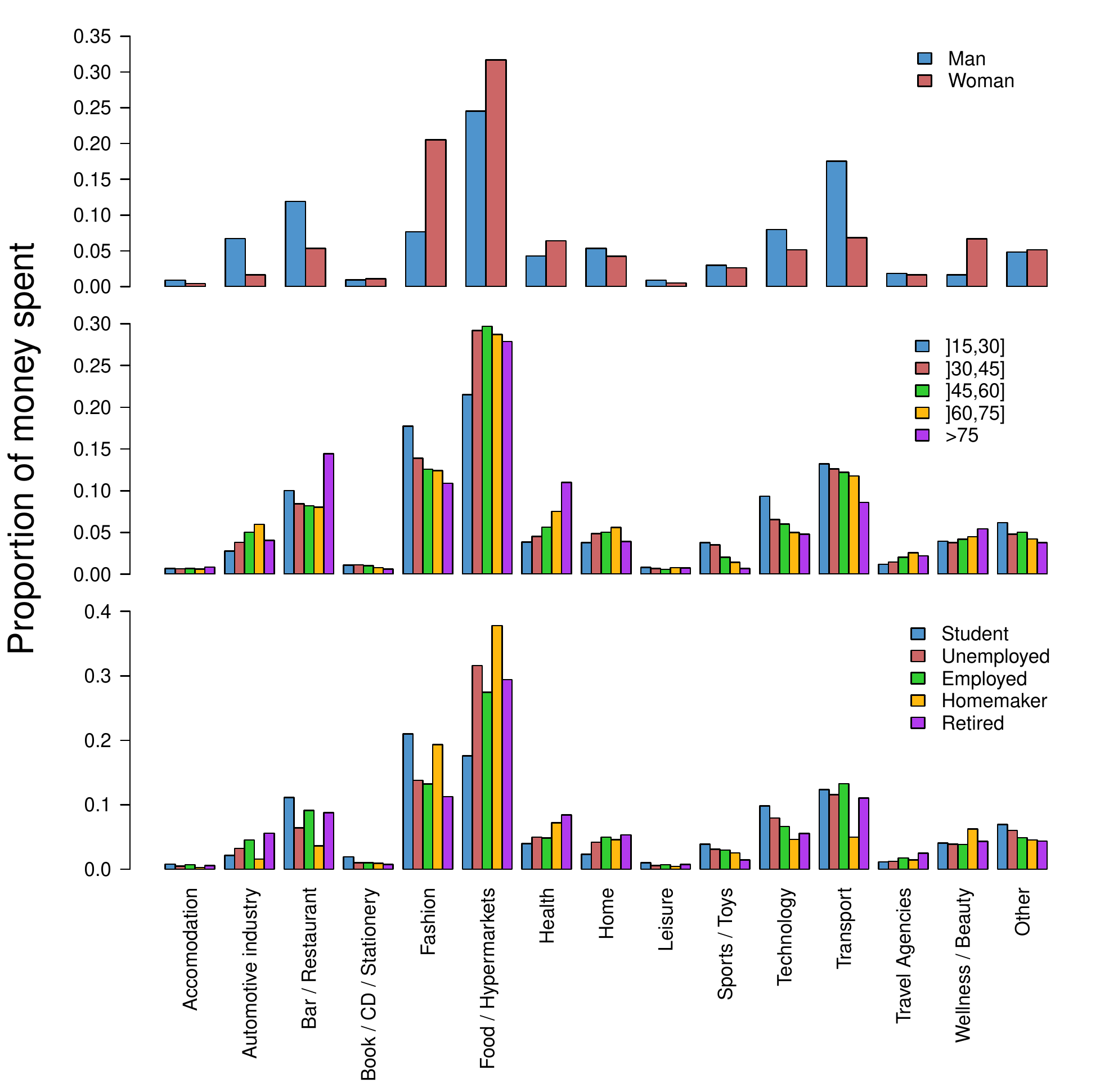}
\caption{\textbf{Average fraction of money spent by an individual according to the business category and his/her sociodemographic characteristics.} From the top to the bottom: gender, age and occupation. \label{Fig3}}
\end{figure*}

\subsection{General features}

In order to have a first look at the data, we plot in Figure \ref{Fig2} some descriptive statistics about individuals according to their sociodemographic characteristics. Figure \ref{Fig2} shows the proportion of individuals according to gender, age and occupation in the dataset and the corresponding fractions as observed in the census \cite{INE2011}. We note an over-representation of men and middle-aged individuals (30-60) in the dataset compared to census data. Moreover, employed people represent about $80\%$ of the individuals, which is two times higher than the proportion of employed people in Spain. Therefore, since the data are not representative of the population, in the rest of the manuscript only indicators and measures normalized by the total number of individuals in each groups will be considered. It is also important to note that the three distributions are not independent, for example, the proportion of individuals according to the age is not the same for student and retired individuals. In the same way, the proportion of individuals according to the occupation is different for men and women. For example, there are more female homemakers than male homemakers. For more details, histograms of the three joint distributions are available in appendix (Figure S1, S2, and S3).  

To highlight differences between individuals having different sociodemographic characteristics, we also plot on Figure \ref{Fig2} the median number of transactions per user, the median amount of money spent per user and the median average amount of money spent per transaction per user. We used the median instead of the average because the distributions exhibits a large number of outliers (see Figure S4, S5 and S6 in appendix for more details). It can be observed that individuals do not spend their money in the same way according to whether they are men or women, young or old and active or inactive. For instance, the number of transactions and the amount of money spent is higher for women than for men and decreases with age. Furthermore, they are also higher for employed persons and homemakers than for unemployed individuals, students and retired people (which is probably related to the age). Inversely, the average  amount of money spent per transaction is higher for men than women and increases with age. 

To investigate the influence of sociodemographics on the way people spend their money, we plot on Figure \ref{Fig3} the average fraction of money spent by an individual according to the business category and his/her sociodemographic characteristics. Since the total amount of money spent in 2011 is different from one individual to another, the distribution has been normalized for each user by the total amount of money he/she spent during the year. Note that the distribution is very different for men and women. Indeed, women spend more money than men in Fashion, Food/Hypermarkets, Health and Wellness/Beauty whereas men spend more money than women in Automotive Industry, Bar/Restaurants, Technology and Transport. We also find that the proportion of money spent in Fashion, Food/Hypermarkets, Sports/Toys, Technology and Transport globally decreases with age. Inversely, the amount of money spent in Automotive Industry, Health, Travel Agencies and Wellness/Beauty increases with age. Finally, the differences between people having different occupation are explored. For instance, students spend more money in Bar/Restaurant, Fashion, Sports/Toys and Technology than others types of occupation. 

Since the proportion of individuals according to the occupation is different for men and women, and in order to take away potential bias, we have studied the average fraction of money spent by an individual according to the business category and his/her sociodemographic characteristics but only for employed individuals. We reach the same conclusions as for the overall sample, see Figure S7 in appendix.

\begin{figure*}
\centering
\includegraphics[scale=0.6]{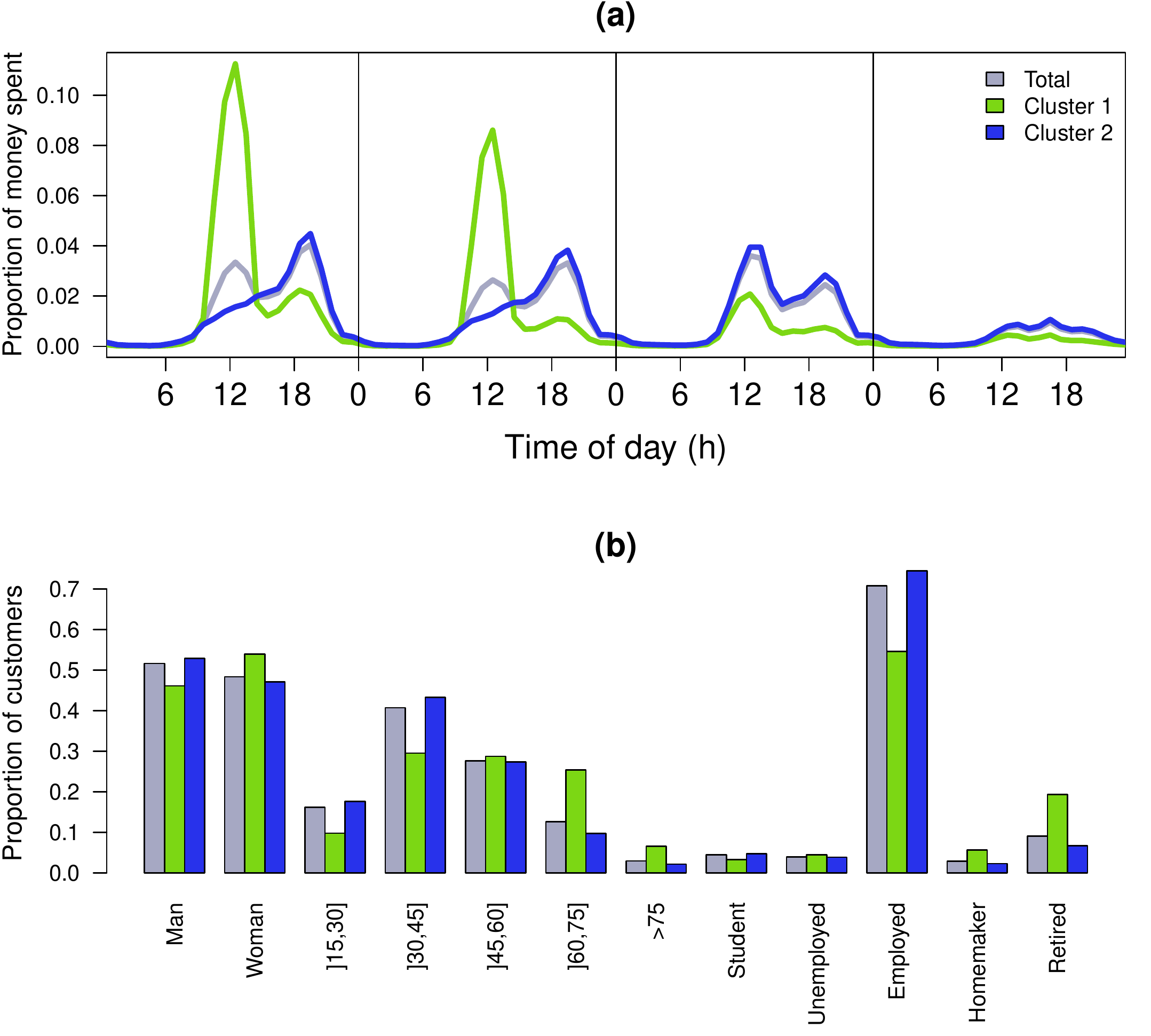}
\caption{\textbf{Time evolution of the amount of money spent.} (a) Average amount spent per day as a function of the hour of the day in total and according to the cluster. From left to right: weekdays (aggregation from Monday to Thursday), Friday, Saturday and Sunday. (b) Proportion of individuals in total and in each cluster according to, from left to right, the gender, the age and the occupation. \label{Fig4}}
\end{figure*}

\subsection{Time evolution of the amount of money spent}

To study how the amount of money spent by BBVA customers changes over time during an average week, the days of the week have been divided into four groups: one, from Monday to Thursday representing a normal working day (hereafter called $WD$) and three more for Friday, Saturday and Sunday (hereafter called $Fri$, $Sat$ and $Sun$). The average amount of money spent per day as a function of the hour of the day is displayed in Figure \ref{Fig4}a (gray curve). Globally, the amount of money spent is significantly higher during the week days, Friday and Saturday than on Sunday. This can be explained by the fact that most of the business were closed on Sunday in Spain in the time that the data was collected. The activity on Sunday takes place between $10am$ and $7pm$ with a small peak around $4pm$. During the week days, Friday and Saturday money is spent between $8am$ and $10pm$. For these days the curves show two peaks, one around noon and another one around $7pm$. It is interesting to note that for the week days and Friday the second peak is higher than the first one whereas the opposite behavior is observed on Saturday. A small peak around $11pm$ corresponding to the nightlife activity is also observed for the three first days.

\begin{figure*}
\centering
\includegraphics[scale=0.4]{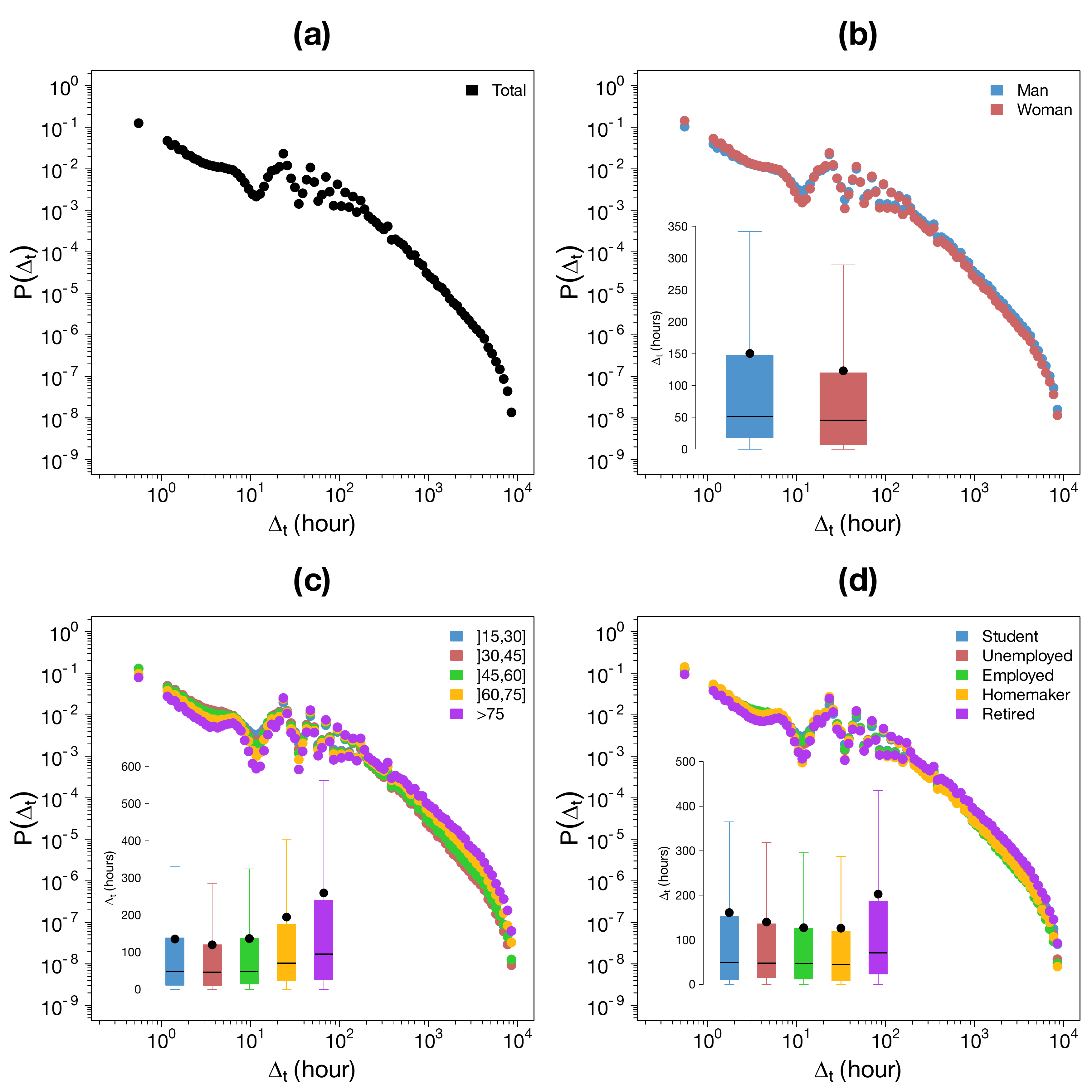}
\caption{\textbf{Inter-event time distribution $P(\Delta_t)$.} (a) Probability density function of $\Delta_t$. (b) -– (d) Probability density function of $\Delta_t$ according to the gender (b), the age (c) and the occupation (d). The insets show the Tukey boxplot of the distributions, the black points represent the average. \label{Fig5}}  
\end{figure*}

To go further in the analysis, a k-means clustering algorithm with Euclidean distance \cite{Hartigan1979} is applied in order to identify clusters naturally present in the data. The purpose is to cluster together individuals exhibiting temporal distribution of money spent. The total amount of money spent in 2011 is different from one individual to another so we have normalized the temporal distribution of money spent for each user by the total amount of money he/she spent in 2011. To choose the number of clusters, we use the pseudo-F statistics which describes the ratio of between-cluster variance to within cluster variance \cite{Calinski1974}. The optimal number of clusters is the one for which the highest pseudo-F value is obtained, in our case we found two opposite clusters (see Figure S8 in appendix for more details). Figure \ref{Fig4}a displays the results of the clustering analysis, we observe an opposition between active and inactive individuals. The first cluster represents one third of the individuals and is characterized by a higher activity during the morning and during weekdays in opposition with the second cluster in which individuals tend to spend more money after $6pm$ and during week end days. It is interesting to note that the first cluster is over-represented by women, old people and homemaker and retired individuals compared to the whole population (Figure \ref{Fig4}b).

\begin{figure*}
\centering
\includegraphics[scale=0.4]{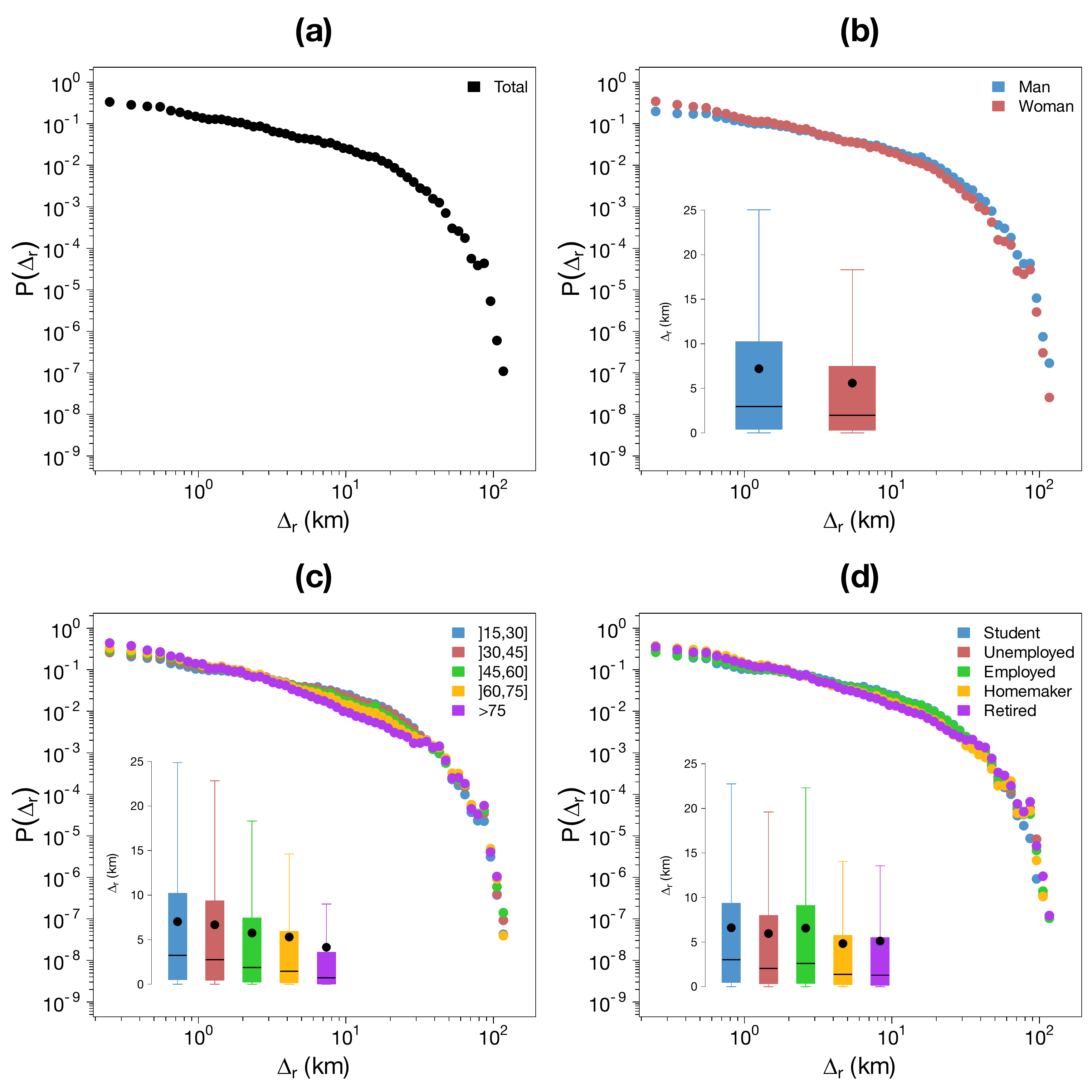}
\caption{\textbf{Distribution of the distance traveled by an individual between two consecutive transactions $P(\Delta_r)$.} (a) Probability density function of $\Delta_r$. (b) -– (d) Probability density function of $\Delta_r$ according to the gender (b), the age (c) and the occupation (d). The insets show the Tukey boxplot of the distributions, the black points represent the average. \label{Fig6}}  
\end{figure*}

\subsection{Mobility patterns}

In order to characterize mobility patterns of each user, we have considered three variables: $\Delta_t$, the time elapsed between two consecutive transactions, $\Delta_r$, the distance traveled between two consecutive transactions, and $r_g$, the radius of gyration \cite{Gonzalez2008}. The radius of gyration is defined as
\begin{equation}
r_g=\sqrt{\frac{1}{n} \sum_{k=1}^{n} (\vec{p_k}-\vec{p_c})^2},
\label{Rg}
\end{equation}
where $\vec{p_k}$ represents the $k^{th}$ position of the user displacements in 2011 and $\vec{p_c}=\frac{1}{n} \sum_{k=1}^{n} \vec{p_k}$  is the center of mass of his/her motions. It is important to note that $r_g$ is defined per user whereas $\Delta_t$ and $\Delta_r$ are computed for each displacement. Although $\Delta_r$ and $r_g$ are related, $\Delta_r$ informs us on the distance traveled by users, which might depend on the frequency at which each person uses its credit card, whereas $r_g$ gives us a more holistic view of how people moves around their centers of mass. To avoid the introduction of bias in the mobility patterns analysis, all the consecutive user's positions geo-located in the province and the distances between them are considered whatever the elapsed time between consecutive transactions.

\begin{figure*}
\centering
\includegraphics[scale=0.4]{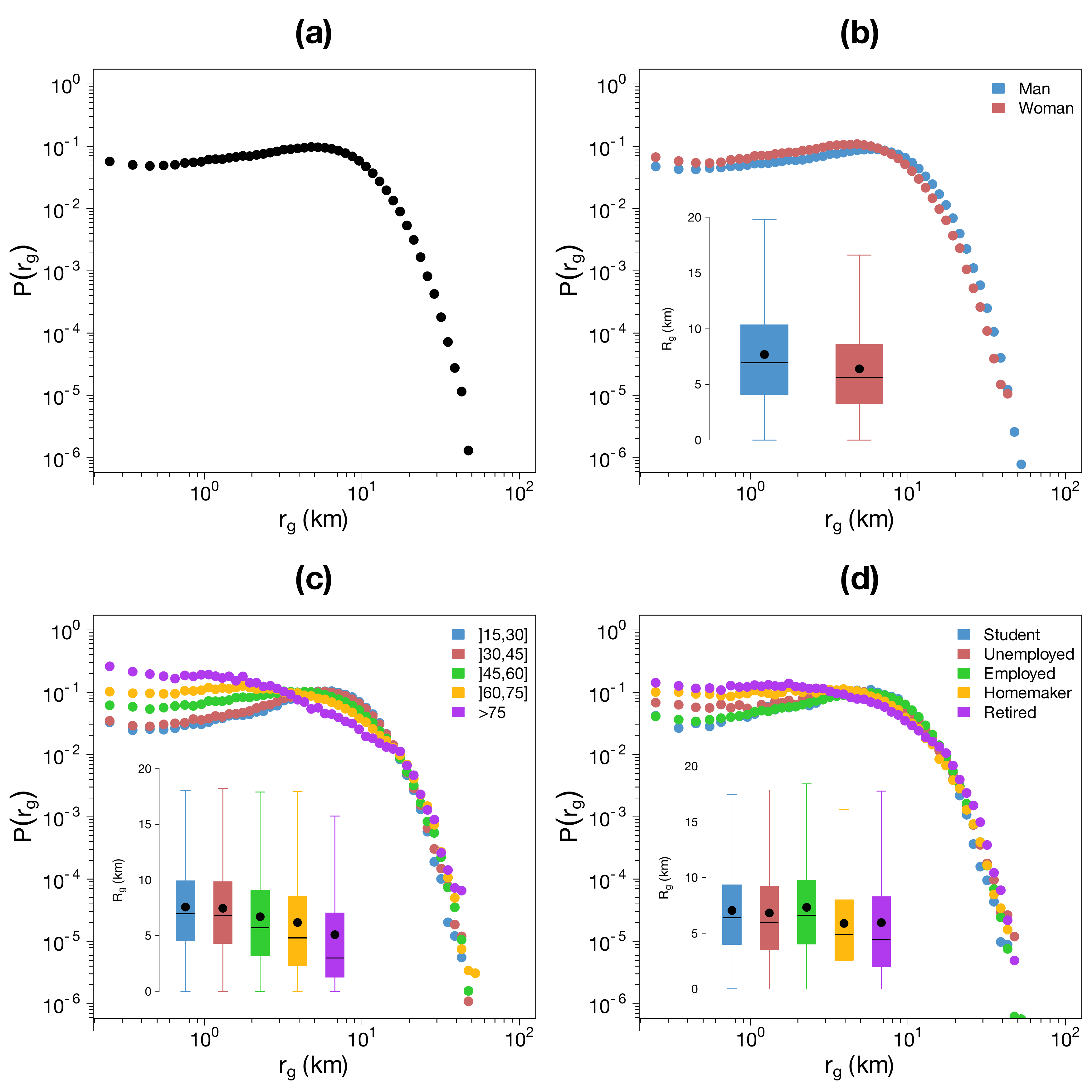}
\caption{\textbf{Distribution of the radius of gyration $P(r_g)$.} (a) Probability density function of $r_g$. (b) -– (d) Probability density function of $r_g$ according to the gender (b), the age (c) and the occupation (d). The insets show the Tukey boxplot of the distributions, the black points represent the average. \label{Fig7}}  
\end{figure*}

Figures \ref{Fig5}a, \ref{Fig6}a and \ref{Fig7}a display the probability density function of the three variables. The distribution of $\Delta_t$ is a decreasing density function exhibiting circadian rhythms. The average and median time between two transaction are, respectively, around $5$ days and $2$ days. The distribution of $\Delta_r$ show two different regimes. First the distribution exhibits a slow decay, and then, beyond $40$ kilometers the distribution is characterized by a rapid decay. This cutoff is introduced by the limited geographical scale of the provinces. The probability density function $P(r_g)$ increases very slowly until reaching a maximum around $6$ kilometers and then the distribution is characterized by a rapid decay.

In this work we have also assessed the influence of sociodemographic characteristics on the individual mobility patterns. The results obtained are plotted on the Figure \ref{Fig5}, \ref{Fig6} and \ref{Fig7}. For each sociodemographic characteristic and each variable, we performed two non-parametric tests to assess the statistical significance of the differences between the different type of users' mobility using the Mann–Whitney U test \cite{Mann1947} to compare the distributions and the Mood's median test \cite{Brown1951} to compare the medians. For both case studies the differences between distributions and medians are always significant (\textit{p-values} lower than $10^{-4}$) except for the difference between radius of gyration of individuals of age between 15 and 30 and those between 30 and 45 in Barcelona. 

Figure \ref{Fig5} displays the inter-event time distribution according to the gender (Figure \ref{Fig5}b), the age (Figure \ref{Fig5}c) and the occupation (Figure \ref{Fig5}d). The average and median inter-event time are higher for men than women and increases with age. They are also higher for unemployed individuals, students and retired people than for employed persons and homemakers. We observe an negative correlation between the time elapsed between two consecutive transactions and the number of transactions per individual described in the first section.

\begin{figure*}
\centering
\includegraphics[width=\linewidth]{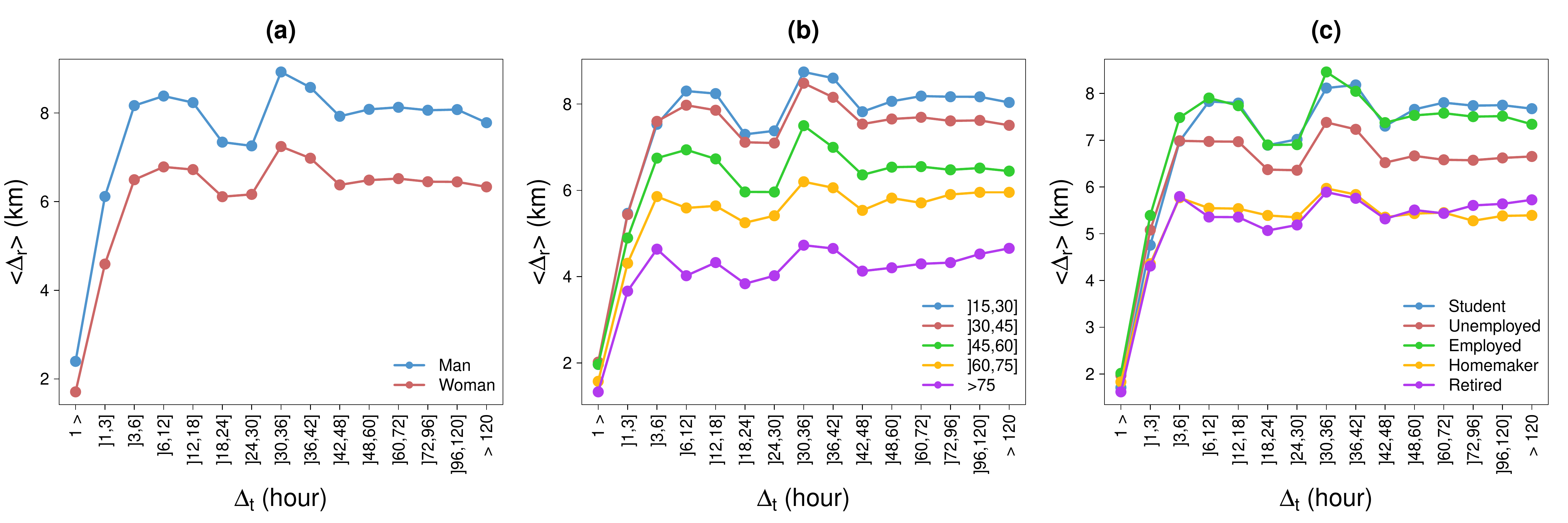}
\caption{\textbf{Average $<\Delta_r>$ value as a function of $\Delta_t$ according to the gender (a), the age (b) and the occupation (c).} \label{Fig8}}  
\end{figure*}

\begin{figure*}
\centering
\includegraphics[width=\linewidth]{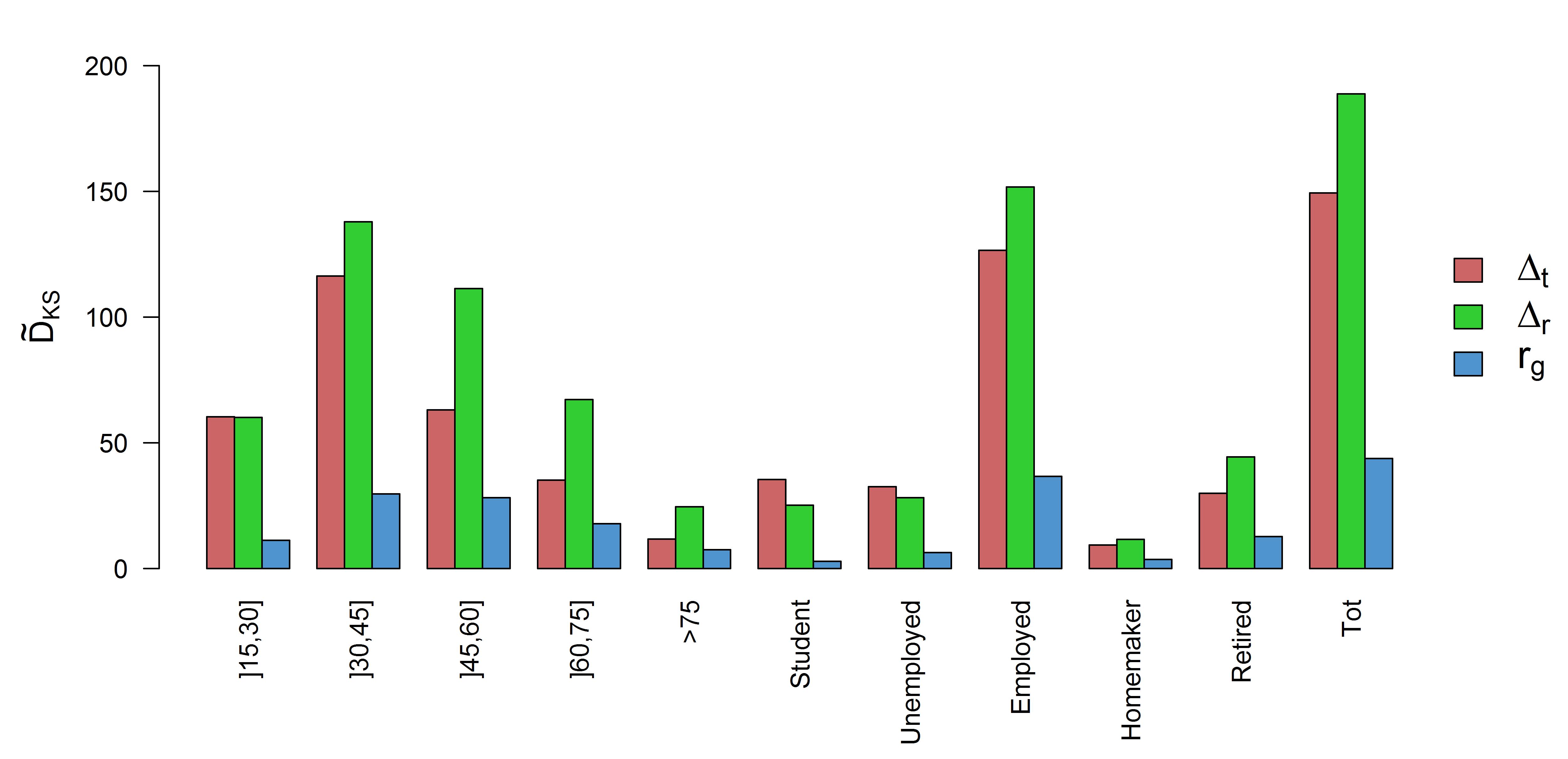}
\caption{\textbf{Kolmogorov-Smirnov distance between men and women's $\Delta_t$ distributions (in red), $\Delta_r$ distributions (in green) and $r_g$ distributions (in blue) according to their sociodemographic characteristics.} \label{Fig9}}
\end{figure*}

The results obtained for $\Delta_r$ and $r_g$ are plotted in Figure \ref{Fig6} and \ref{Fig7}, respectively. Based on these results, one can understand that, depending on his/her sociodemographic characteristics, an individual can travel short or long distances and stays more or less close to his/her center of mass. Three main differences are observed. First, women travel shorter distances than men and their trajectory stays closer to their center of mass. Second, the average distance traveled between two consecutive positions and the radius of gyration decrease with age. Finally, an opposition between active and inactive individual is highlighted. Indeed, retired, homemaker and, to a lesser extent, unemployed individuals travel shorter distances and stay closer to the center of mass than other people. 

As previously mentioned, the distance traveled by an individual between two consecutive transactions might depend on the frequency at which an individual uses his/her credit card, and therefore, the differences between people observed for $\Delta_r$ could be a consequence of the differences observed for $\Delta_t$. Although the same conclusion are reached for the radius of gyration, which does not depend on the frequency at which someone uses his/her credit card, it could be interesting to study how the average value of $\Delta_r$ evolves as a function of $\Delta_t$ according to the individual's sociodemographic characteristics. We can observe in Figure \ref{Fig8} that the differences between the different types of individuals in terms of distances traveled always exist whatever the time elapsed between two consecutive transactions. It is also worth noting that the value of $<\Delta_r>$ is not completely independent of $\Delta_t$. Obviously, for small values of $\Delta_t$ ($\Delta_t < 6$) the value of $<\Delta_r>$ increases with the value of $\Delta_t$ due to physical constraints but we can also note a valley for $\Delta_t \in \, ]18,30]$ followed by a peak for $\Delta_t \in \, ]30,42]$. This phenomenon seems to be more pronounced for active people than for inactive people, possibly reflecting the home-to-work/school commuting.

Among all these comparisons, discrepancy in mobility between men and women is the most challenging. In order to verify that this difference is significant and it is not related to other sociodemographic variables, the Kolmogorov-Smirnov (KS) distance between men and women's $\Delta_t$, $\Delta_r$ and $r_g$ distributions are computed (Figure \ref{Fig9}). The Kolmogorov-Smirnov (KS) distance between two empirical probability distributions $X$ and $Y$ is defined as

\begin{equation}
D_{KS}= \sup_x |F_X(x) - F_Y(x)|,
\label{DKS}
\end{equation}

where $F_X$ and $F_Y$ are the empirical cumulative distribution function of $X$ and $Y$ respectively. Since, the sample size of both distributions may vary from one sociodemographic variable to another we need to normalize $D_{KS}$ according to the sample sizes,

\begin{equation}
\tilde{D}_{KS}= \sqrt{\frac{n_X n_Y}{n_X+n_Y}} \sup_x |F_X(x) - F_Y(x)|,
\label{DKS2}
\end{equation}

where $n_X$ and $n_Y$ represent the sample sizes of $X$ and $Y$, respectively. This allows for a direct comparisons of the Kolmogorov-Smirnov distances. Moreover, using this normalization, the null hypothesis that the two data samples come from the same distribution is rejected at level $0.001$ if $\tilde{D}_{KS} > 1.95$.

First, we observe that a significant difference between men and women appears whatever the sociodemographic characteristic of the population is filtered out (i.e. $\tilde{D}_{KS}$ is always higher than 1.95), which means that on average, women have an inter-event time lower than men and men do longer journeys than women. Second, one can observe that this gendered difference is more important for middle age individual than for young and old people, but also that is more pronounced for employed people.

\begin{figure*}
\centering
\includegraphics[scale=0.6]{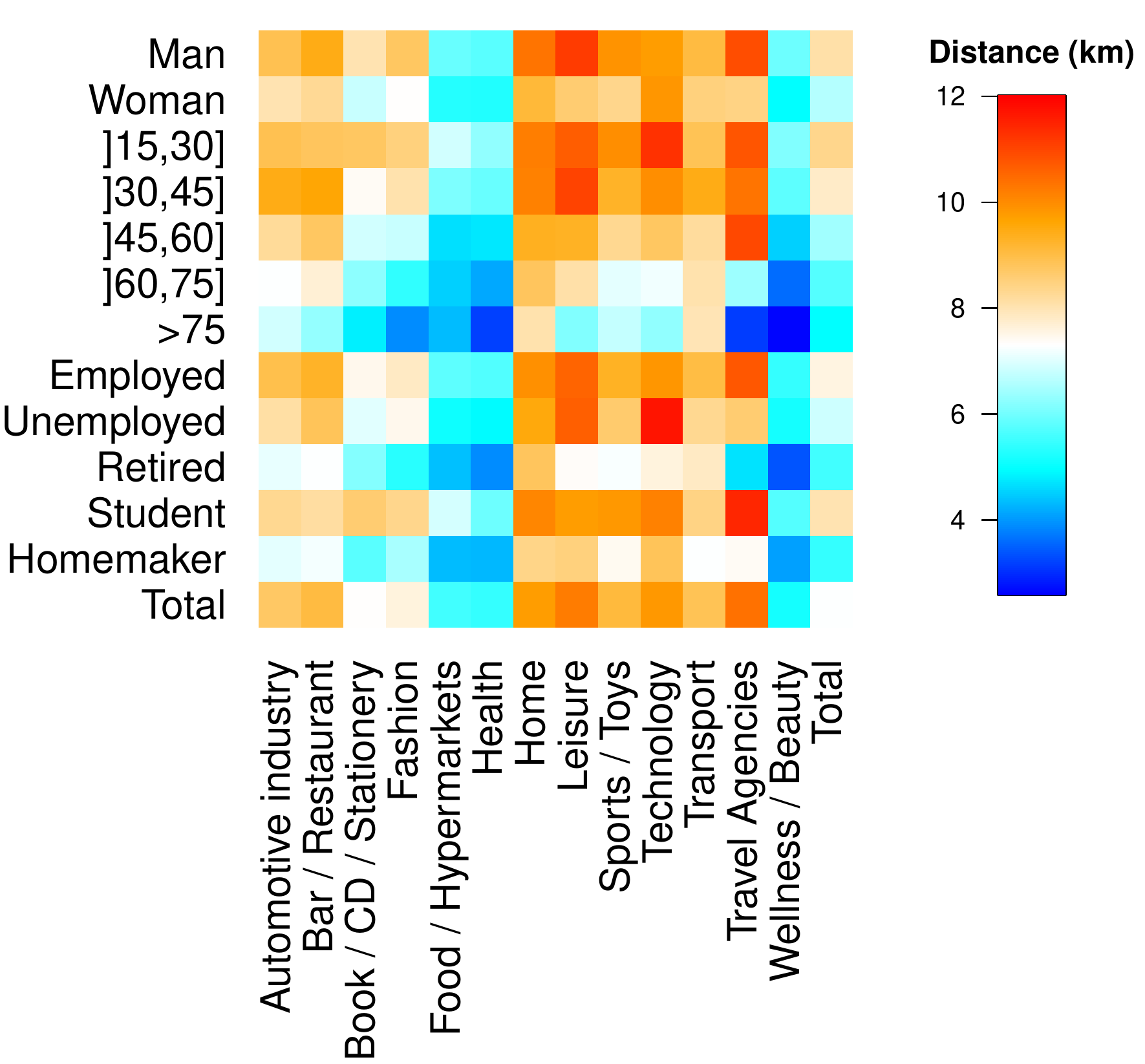}
\caption{\textbf{Average distance between individual’s residence and business according to sociodemographics and business category.} Distances are expressed in kilometer and are computed using the Haversine distance between the latitude and longitude coordinate of the centroid of the postcode of residence and the business' latitude and longitude coordinates for each transaction. \label{Fig10}}
\end{figure*} 

\begin{figure*}
\centering
\includegraphics[width=\linewidth]{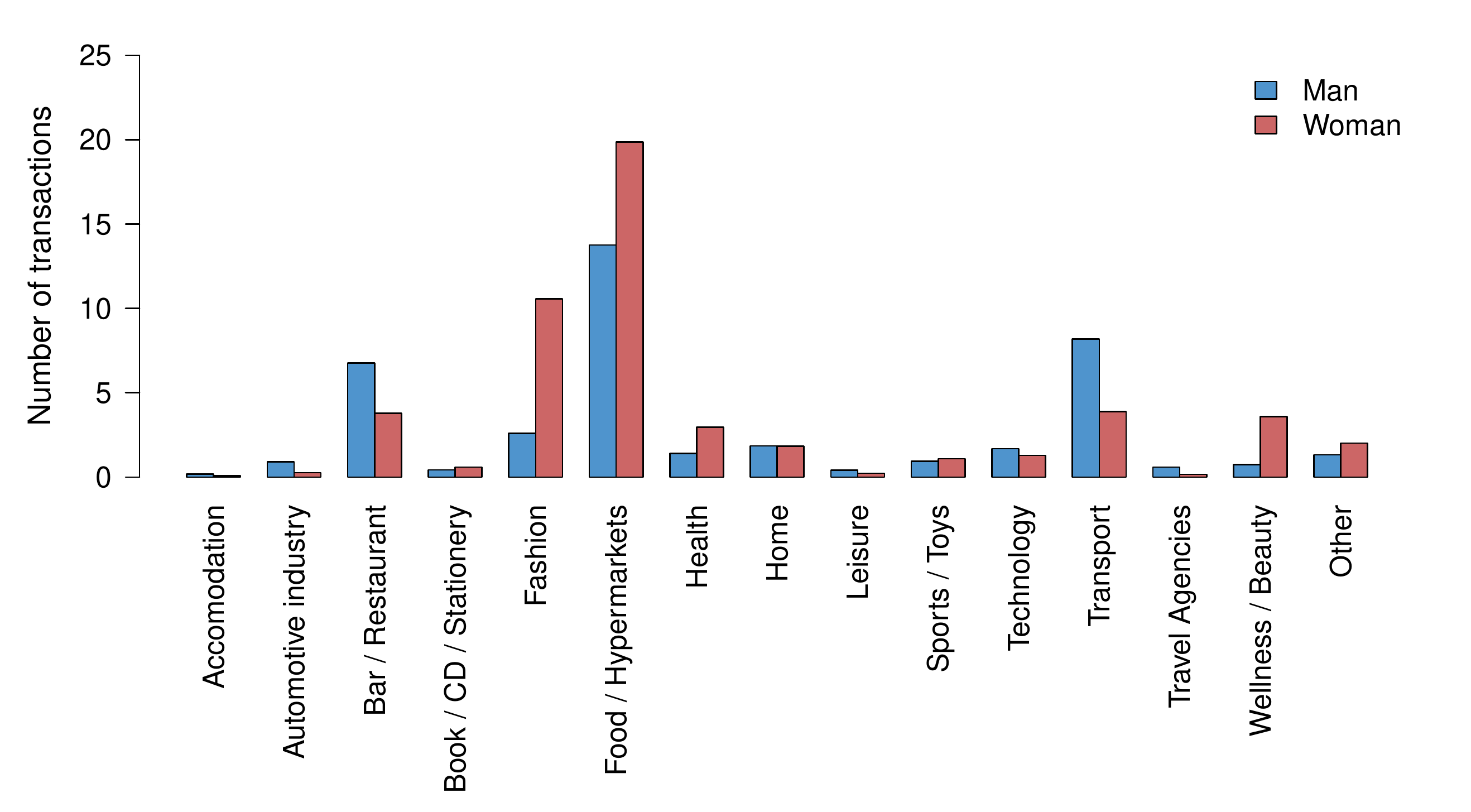}
\caption{\textbf{Average number of transactions according to the gender and the business category.} \label{Fig11}}
\end{figure*}

To go further, we have studied too the influence of the individual’s sociodemographic characteristics and the business category on the distance traveled between home and business. To do so, we computed for each transaction the distance between the individual's place of residence and the business. As residence location, we use  the centroid of the individual's postcode of residence. Finally, these distances were averaged according to individual and business type. These average distances can be observed in Figure \ref{Fig10}. First, we observe that the same differences between type of individuals as the ones highlighted previously are obtained whatever the business category. For each business category, the distance between home and business is globally higher for men than women, it decreases with age and it is higher for employed and student than for the other occupation categories. Although, the average distance between home and business changes according to the category of business. Indeed, distances between home and businesses belonging to the categories Food/Hypermarkets, Health, Wellness/Beauty and Book/CD/Stationery are lower than for the other categories. It is interesting to note that these business category are also the type of business in which the number of transactions is higher for women than for men (Figure \ref{Fig11}). This partially explains why women travel shorter distance than men to go shopping.

\section{DISCUSSION}

In summary, we have shown in this study that it is possible to use information provided by credit card data to assess the influence of sociodemographic characteristics on the way people move and spend their money. We highlighted differences in consumption habits and mobility patterns of bank customers according to their gender, age and occupation. First, we shown that according to the business type the fraction of money spent can be very different from one individual to another. In particular, women tend to spend more money in Fashion, Food/Hypermarkets, Health and Wellness/Beauty than men whereas men spend more money than women in Automotive Industry, Bar/Restaurants, Technology and Transport. We have also studied the time evolution of the amount of money spent along the week according to the individual's sociodemographic characteristics. An opposition between two types of individuals has been identified. The temporal distribution of money spent by the first type of individuals which is over-represented by inactive people is characterized by a higher activity during the morning and during weekdays in opposition with the second type of individuals more active after working hours and during week end days. Then, we investigated the properties of people mobility patterns using three variables: the time elapsed between two consecutive transactions, the distance traveled by an individual between two consecutive transactions and the radius of gyration. Three main differences between groups of people were identified: differences between men and women, young and old people and active and inactive individuals. In the three cases, people of the first group (men, young people and active people) travel shorter distances and their trajectory stays closer to their center of mass than individuals of the second groups (women, old individual and inactive people). 

Among all the differences emphasized in this paper the one between men and women is the most difficult to explain. In all the comparisons we have carefully checked that this difference was not related to other sociodemographic variables and it was not the case. It could be interesting to verify whether this difference is related to other social characteristics such as the number of children for example. Indeed, the fact that the difference in terms of mobility patterns between men and women is less pronounced for old people and students may reflect that women with children move differently than women without children. While further data is required to assess whether these differences between individuals are universal, i.e., to which extend they are specific or not to urban areas or the cities of the country analyzed, our results point toward the possibility that mobility may display significant differences for different types of individuals. 

\vspace*{0.5cm}
\section{ACKNOWLEDGEMENTS}
Partial financial support has been received from the Spanish Ministry of Economy (MINECO) and FEDER (EU) under projects MODASS (FIS2011-24785) and INTENSE@COSYP (FIS2012-30634),  and from the EU Commission through projects EUNOIA, LASAGNE and INSIGHT. The work of ML has been funded under the PD/004/2013 project, from the Conselleria de Educación, Cultura y Universidades of the Government of the Balearic Islands and from the European Social Fund through the Balearic Islands ESF operational program for 2013-2017. JJR acknowledges funding from the Ram\'on y Cajal program of MINECO.

\newpage
\clearpage
\newpage

\makeatletter
\renewcommand{\fnum@figure}{\small\textbf{\figurename~S\thefigure}}
\makeatother
\setcounter{figure}{0}

\section*{APPENDIX}

\onecolumngrid

\begin{figure}[!hb]
\centering
\includegraphics[scale=0.65]{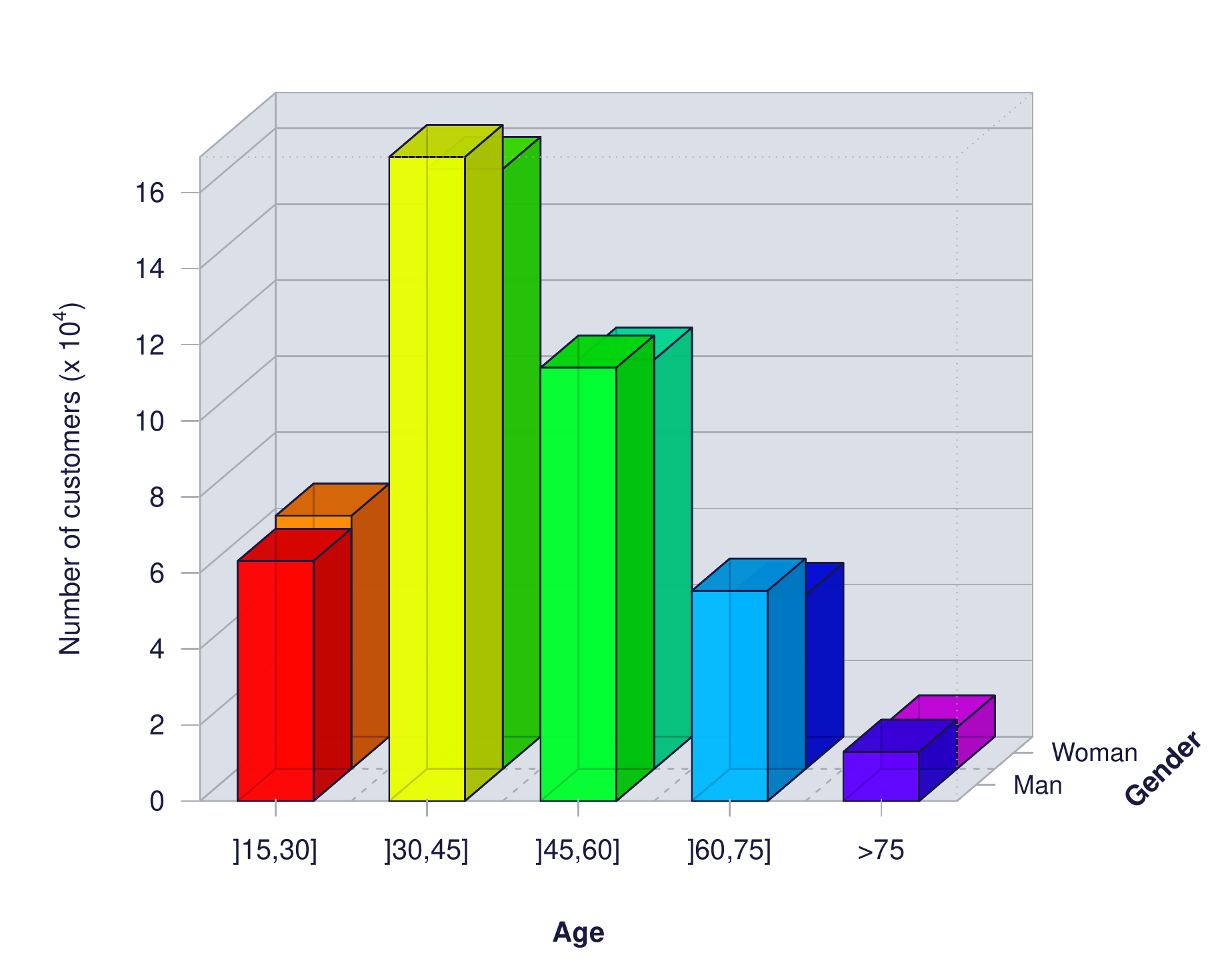}
\caption{\textbf{Histogram of the joint distribution of individuals according to the gender and the age.} \label{FigS1}}
\end{figure} 

\begin{figure}[!hb]
\centering
\includegraphics[scale=0.65]{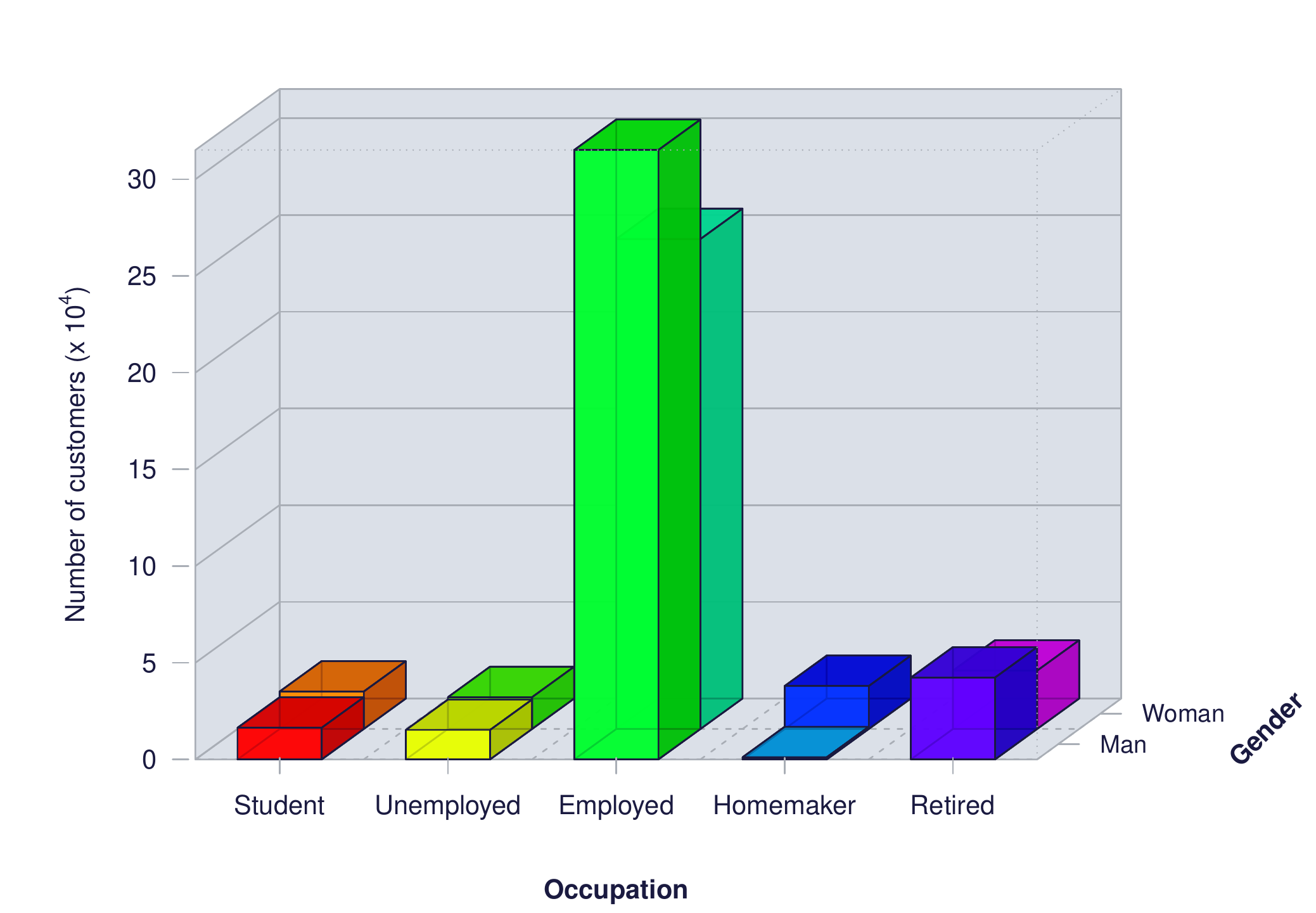}
\caption{\textbf{Histogram of the joint distribution of individuals according to the gender and the occupation.} \label{FigS2}}
\end{figure} 

\newpage
\vspace*{5cm}
\begin{figure}
\centering
\includegraphics[width=\linewidth]{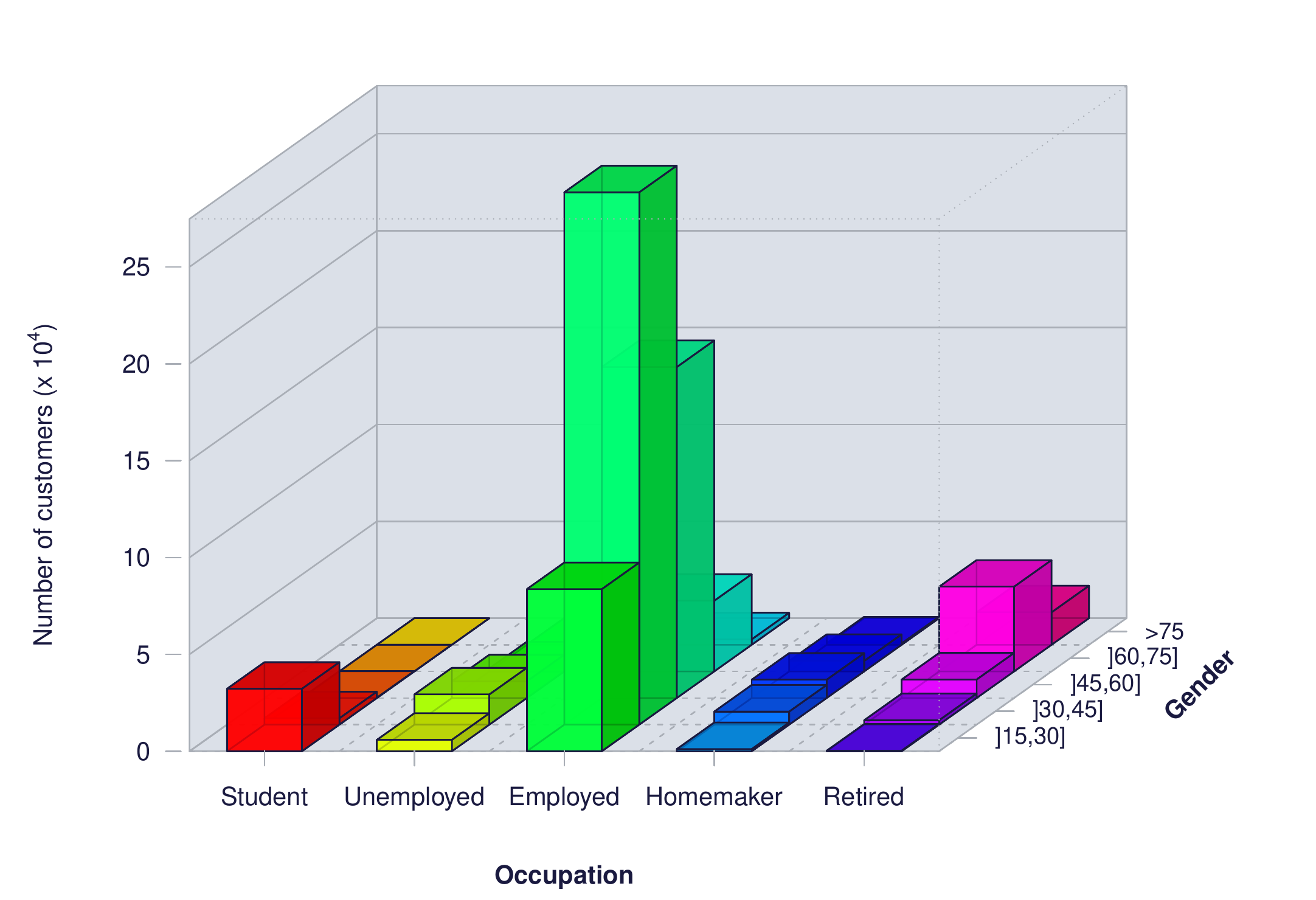}
\caption{\textbf{Histogram of the joint distribution of individuals according to the age and the occupation.} \label{FigS3}}
\end{figure}

\begin{figure*}
\centering
\includegraphics[width=\linewidth]{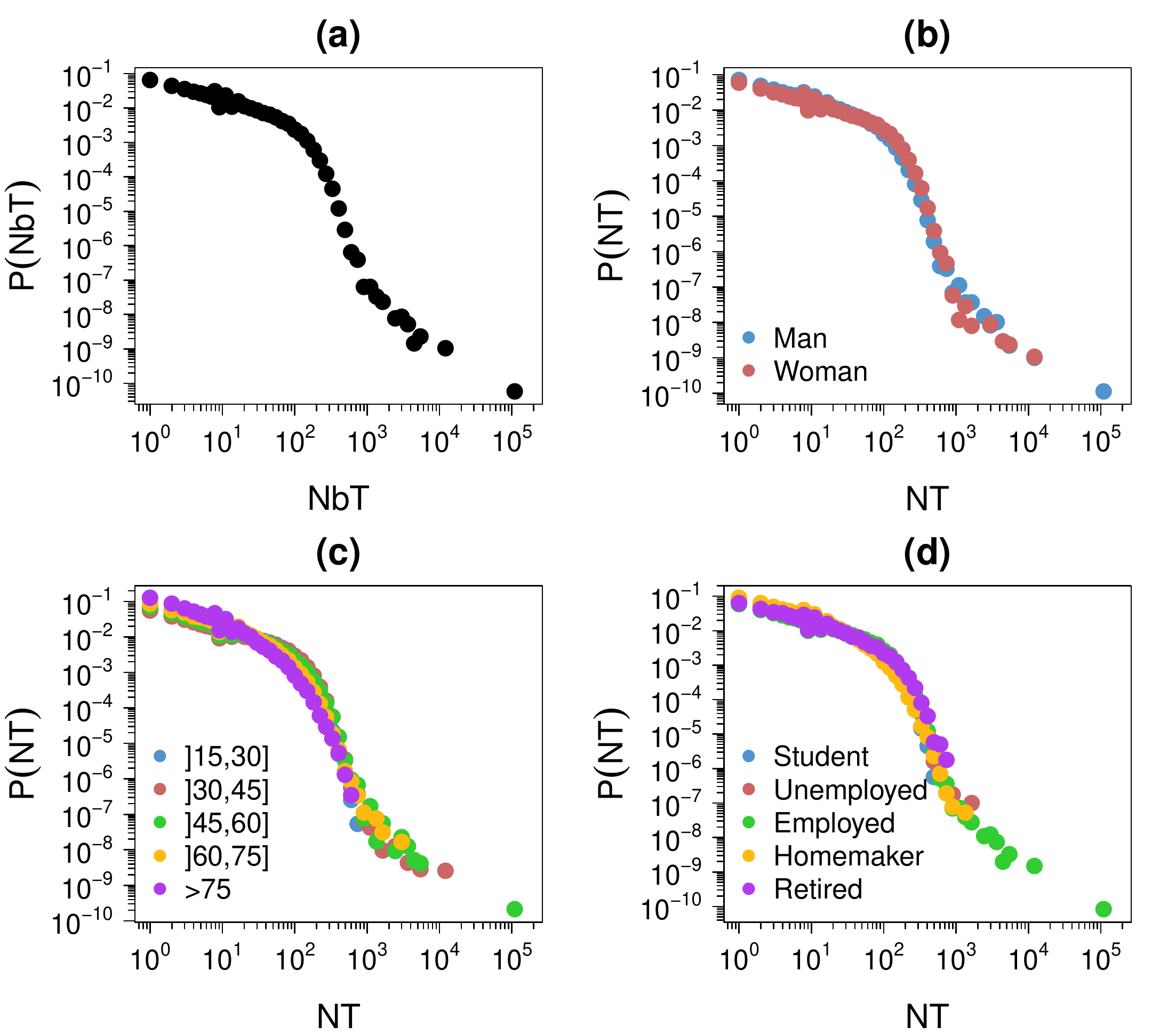}
\caption{\textbf{Probability density function of the number of transactions per individual (a), according to the gender (b), the age (c) and the occupation (d).} \label{FigS4}}
\end{figure*}

\begin{figure*}
\centering
\includegraphics[width=\linewidth]{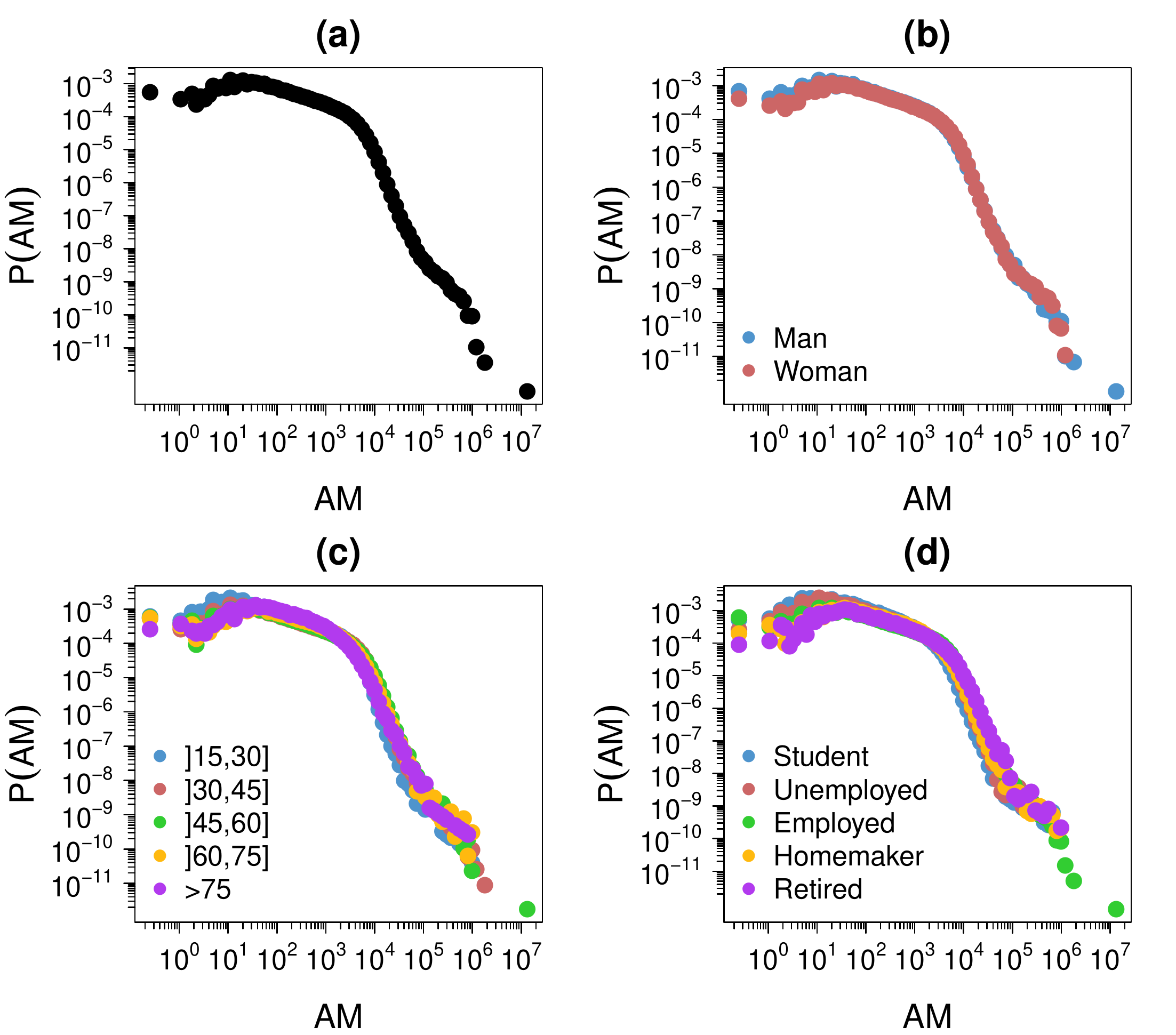}
\caption{\textbf{Probability density function of the amount of money spent in 2011 per individual (a), according to the gender (b), the age (c) and the occupation (d).} \label{FigS5}}
\end{figure*}

\begin{figure*}
\centering
\includegraphics[width=\linewidth]{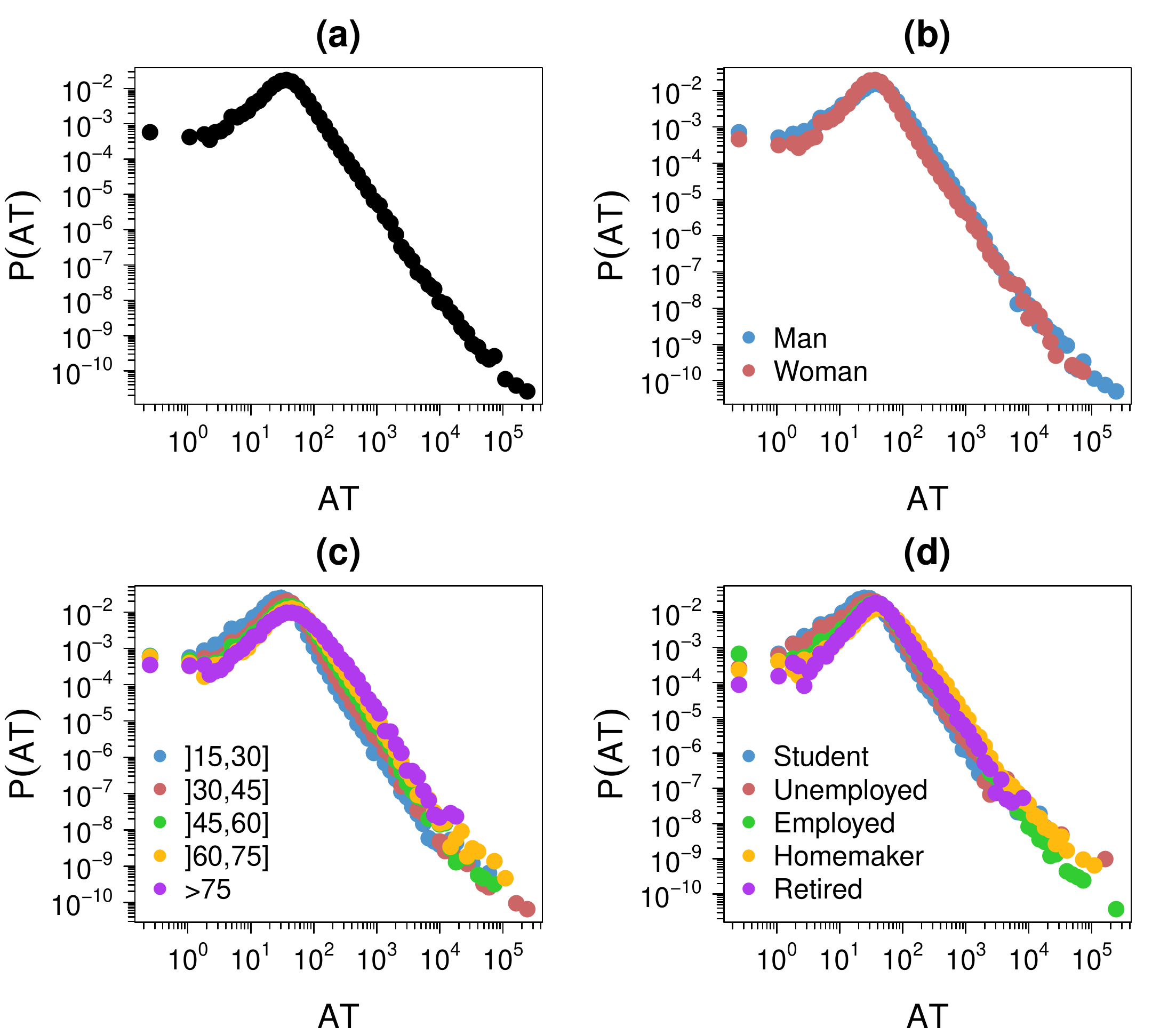}
\caption{\textbf{Probability density function of the average amount of money spent per transaction and per individual (a), according to the gender (b), the age (c) and the occupation (d).} \label{FigS6}}
\end{figure*}

\begin{figure*}
\centering
\includegraphics[width=\linewidth]{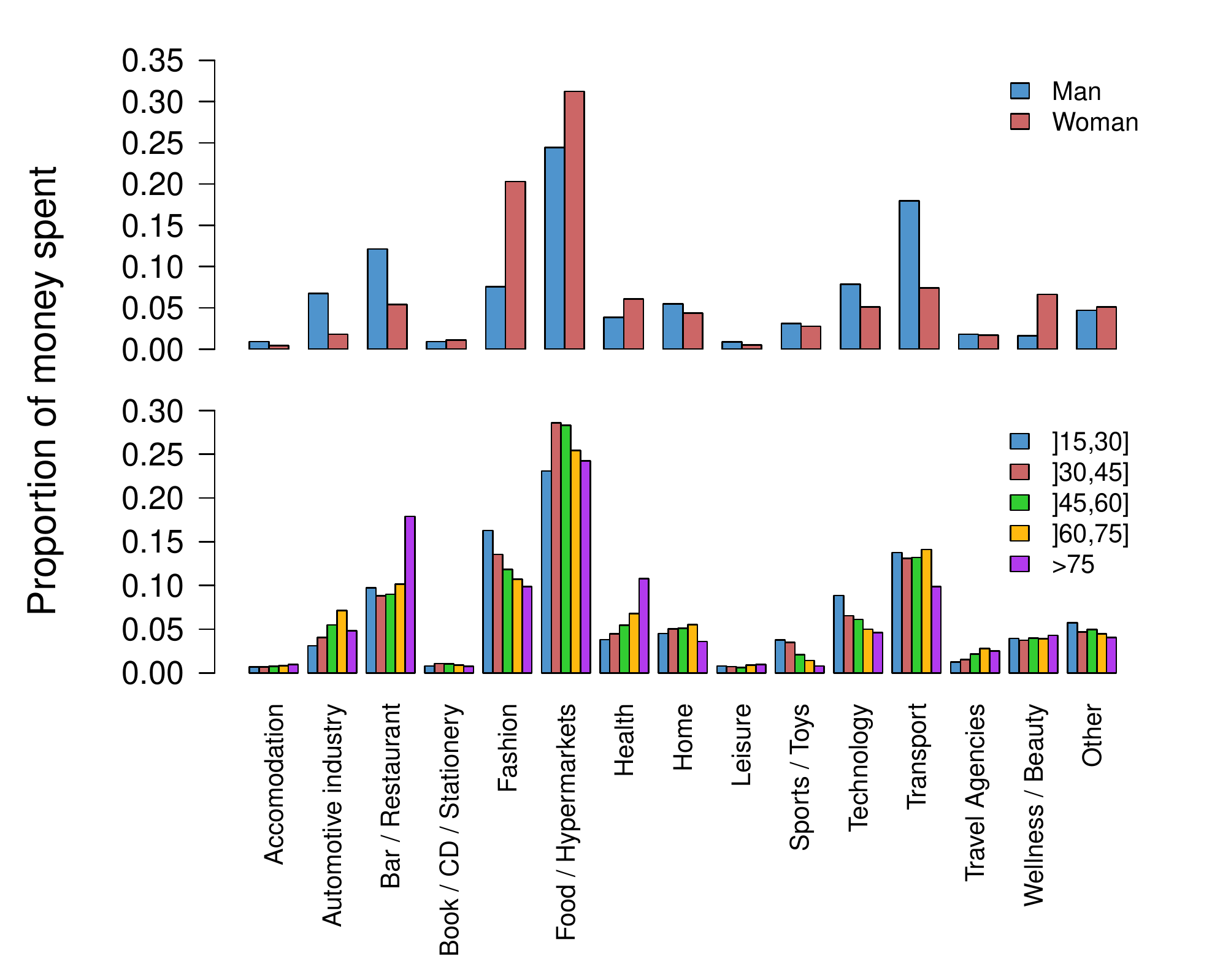}
\caption{\textbf{Average fraction of money spent by an employed individual according to the business category and to his/her gender and age.}  \label{FigS7}}
\end{figure*}

\begin{figure*}
\centering
\includegraphics[scale=0.5]{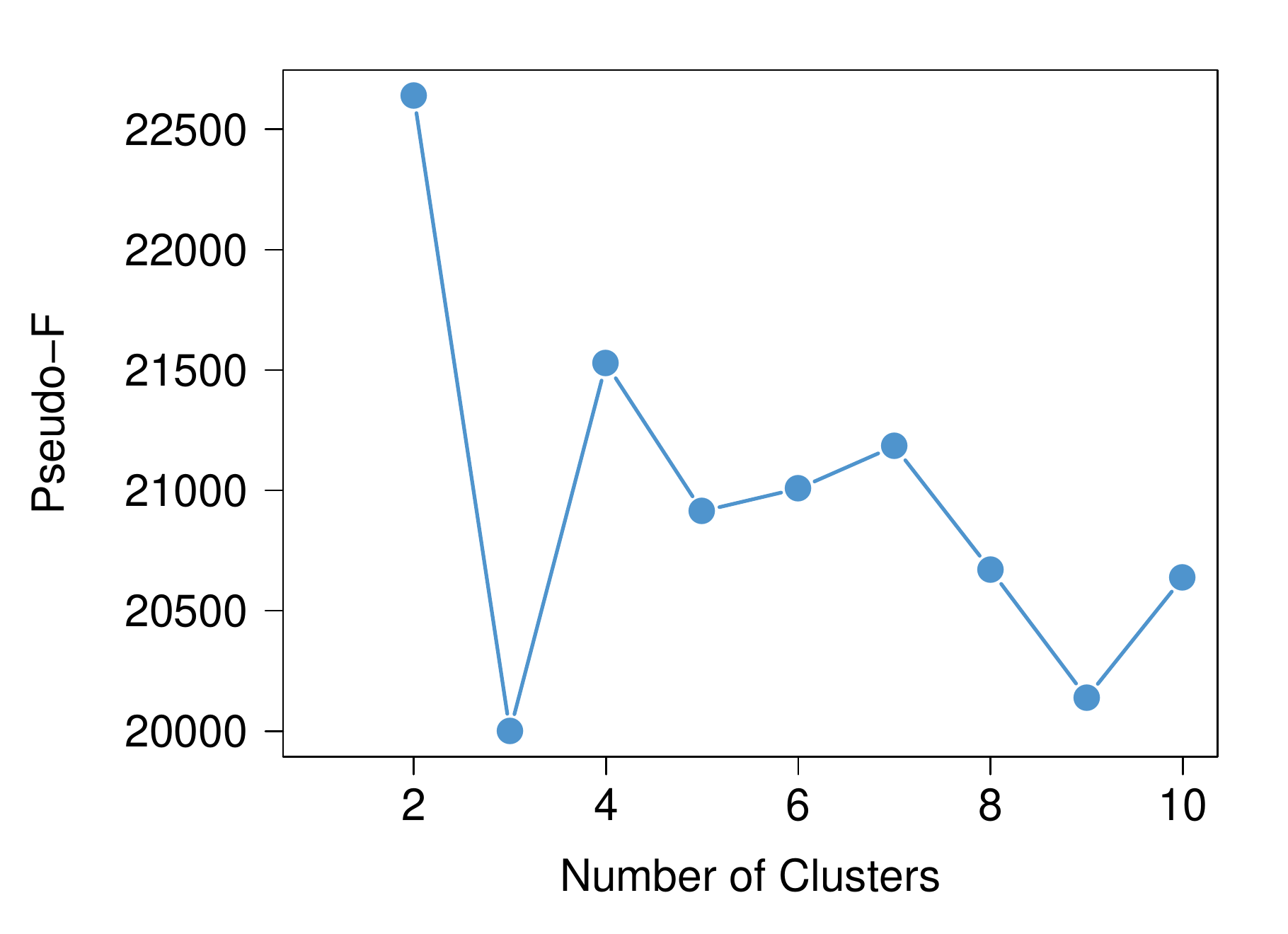}
\caption{\textbf{Pseudo-F as a function of the number of clusters.} K-means clustering algorithm with Euclidean distance applied on the normalized distributions of money spent according to the hour of the day. \label{FigS8}}
\end{figure*}

\begin{figure*}
\centering
\includegraphics[width=\linewidth]{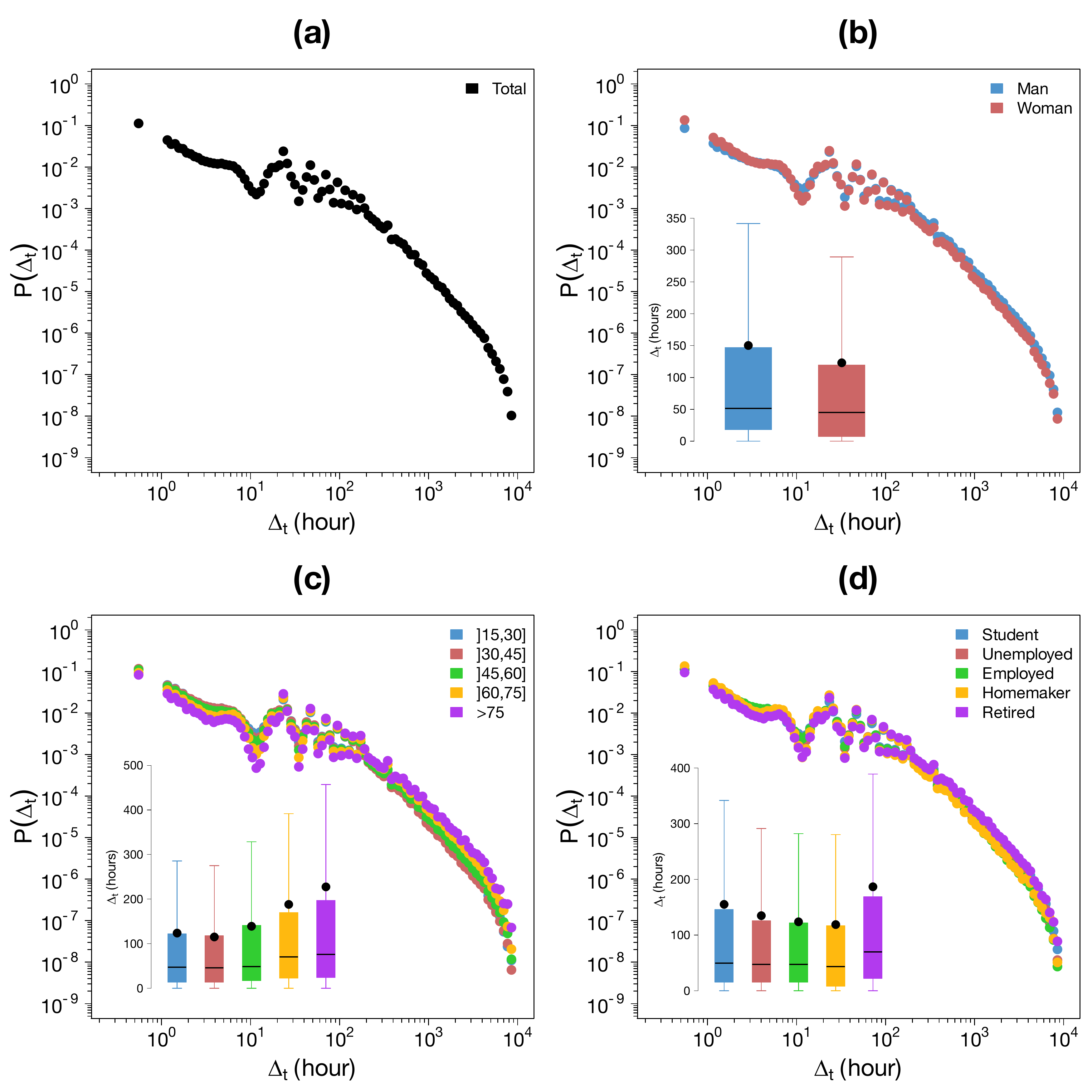}
\caption{\textbf{Inter-event time distribution $P(\Delta_t)$.} (a) Probability density function of $\Delta_t$. (b) -– (d) Probability density function of $\Delta_t$ according to the gender (b), the age (c) and the occupation (d). The insets show the Tukey boxplot of the distributions, the black points represent the average. \label{FigS9}}
\end{figure*}

\begin{figure*}
\centering
\includegraphics[width=\linewidth]{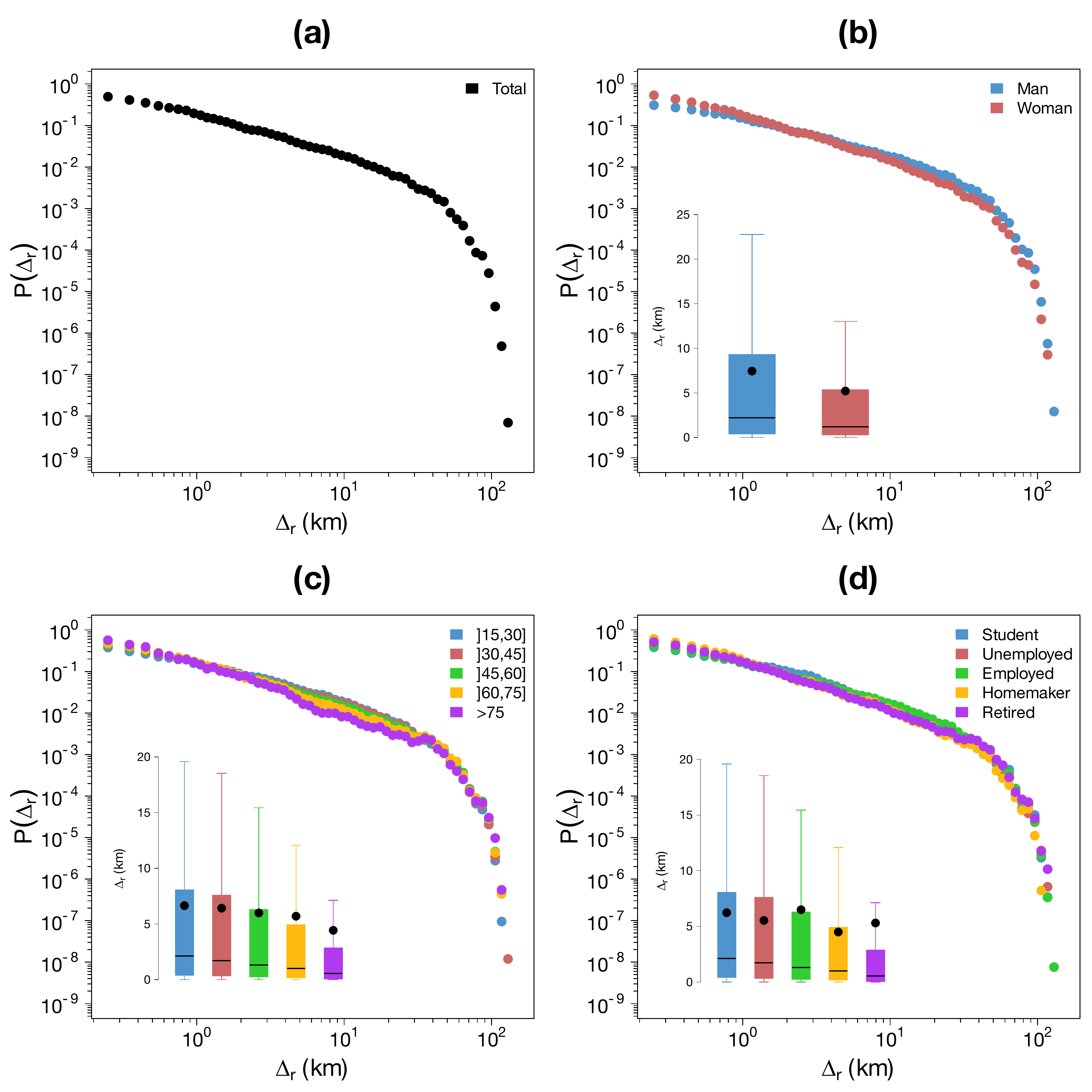}
\caption{\textbf{Distribution of the distance traveled by a customer between two consecutive transactions $P(\Delta_r)$.} (a) Probability density function of $\Delta_r$. (b) -– (d) Probability density function of $\Delta_r$ according to the gender (b), the age (c) and the occupation (d). The insets show the Tukey boxplot of the distributions, the black points represent the average. \label{FigS10}}
\end{figure*}

\begin{figure*}
\centering
\includegraphics[width=\linewidth]{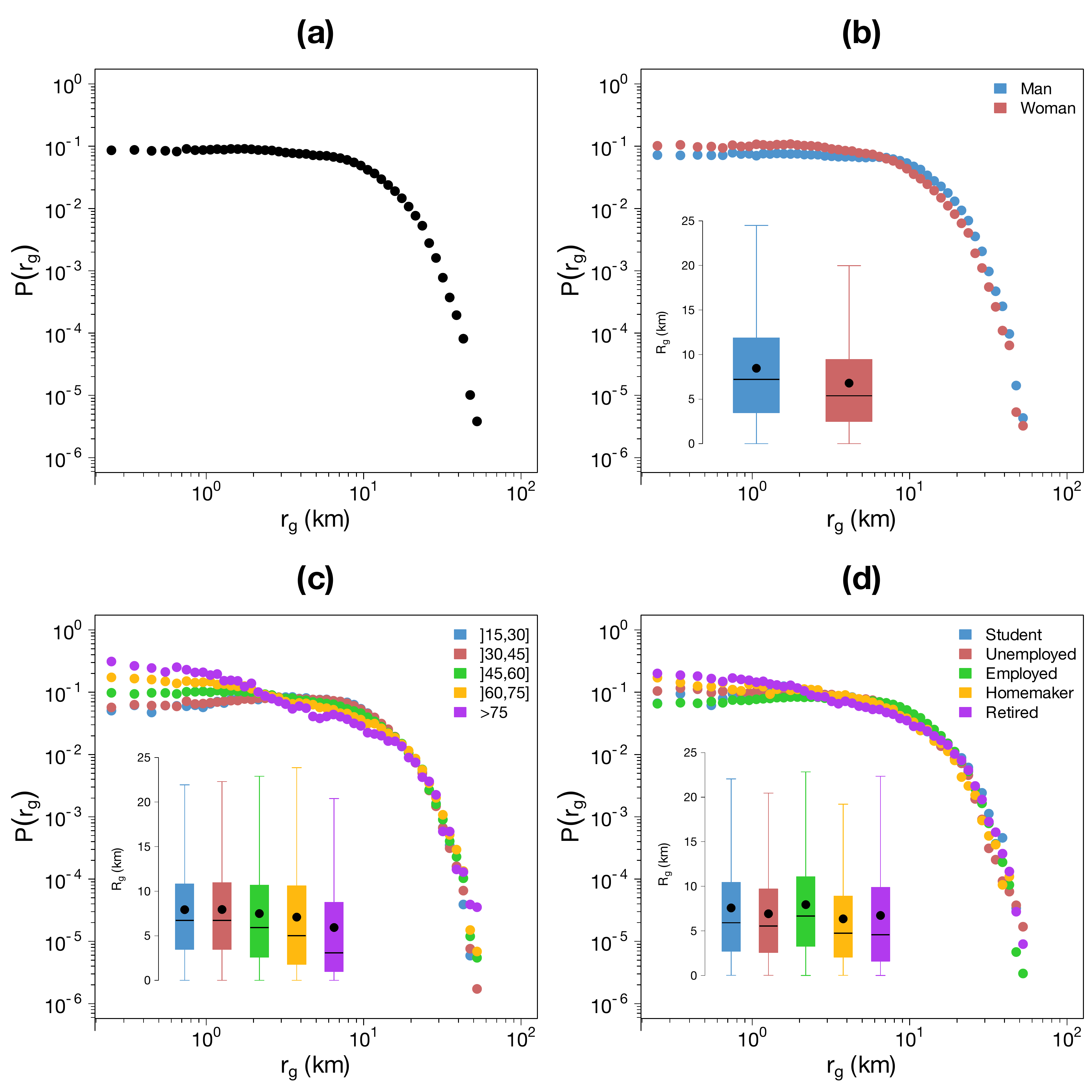}
\caption{\textbf{Distribution of the radius of gyration $P(r_g)$.} (a) Probability density function of $r_g$. (b) -– (d) Probability density function of $r_g$ according to the gender (b), the age (c) and the occupation (d). The insets show the Tukey boxplot of the distributions, the black points represent the average. \label{FigS11}}
\end{figure*}

\begin{figure*}
\centering
\includegraphics[width=\linewidth]{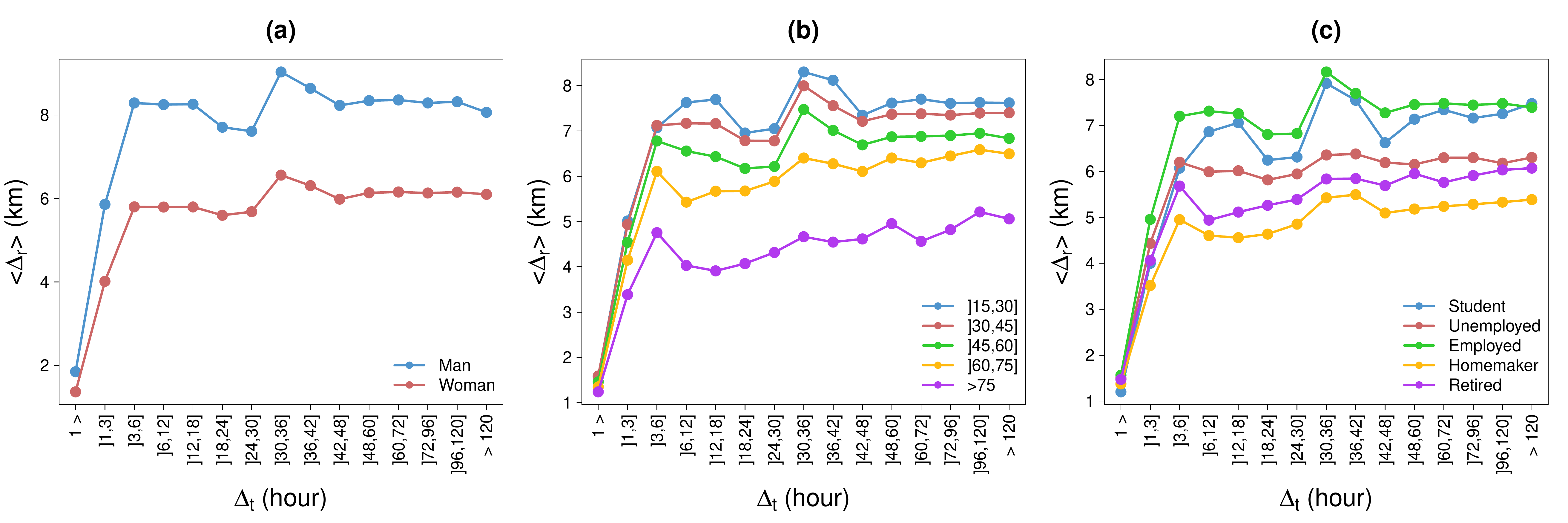}
\caption{\textbf{Average $<r_g>$ value as a function of $\Delta_t$ according to the gender (a), the age (b) and the occupation (c).} \label{FigS12}}
\end{figure*}

\begin{figure*}
\centering
\includegraphics[width=\linewidth]{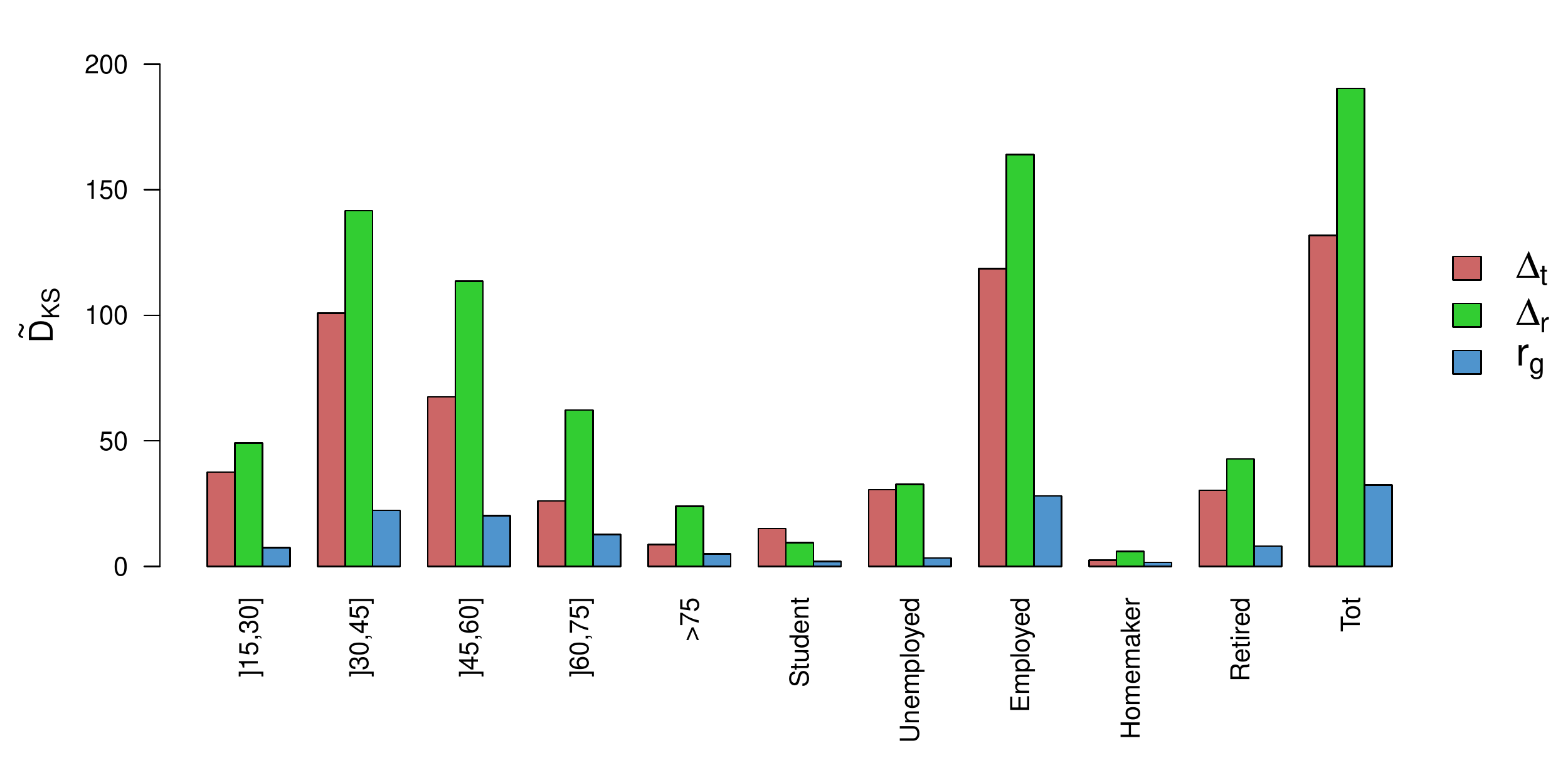}
\caption{\textbf{Kolmogorov-Smirnov distance between men and women's $\Delta_t$ distributions (in red), $\Delta_r$ distributions (in green) and $r_g$ distributions (in blue) according to their demographic characteristics.} \label{FigS13}}
\end{figure*}

\begin{figure*}
\centering
\includegraphics[scale=0.7]{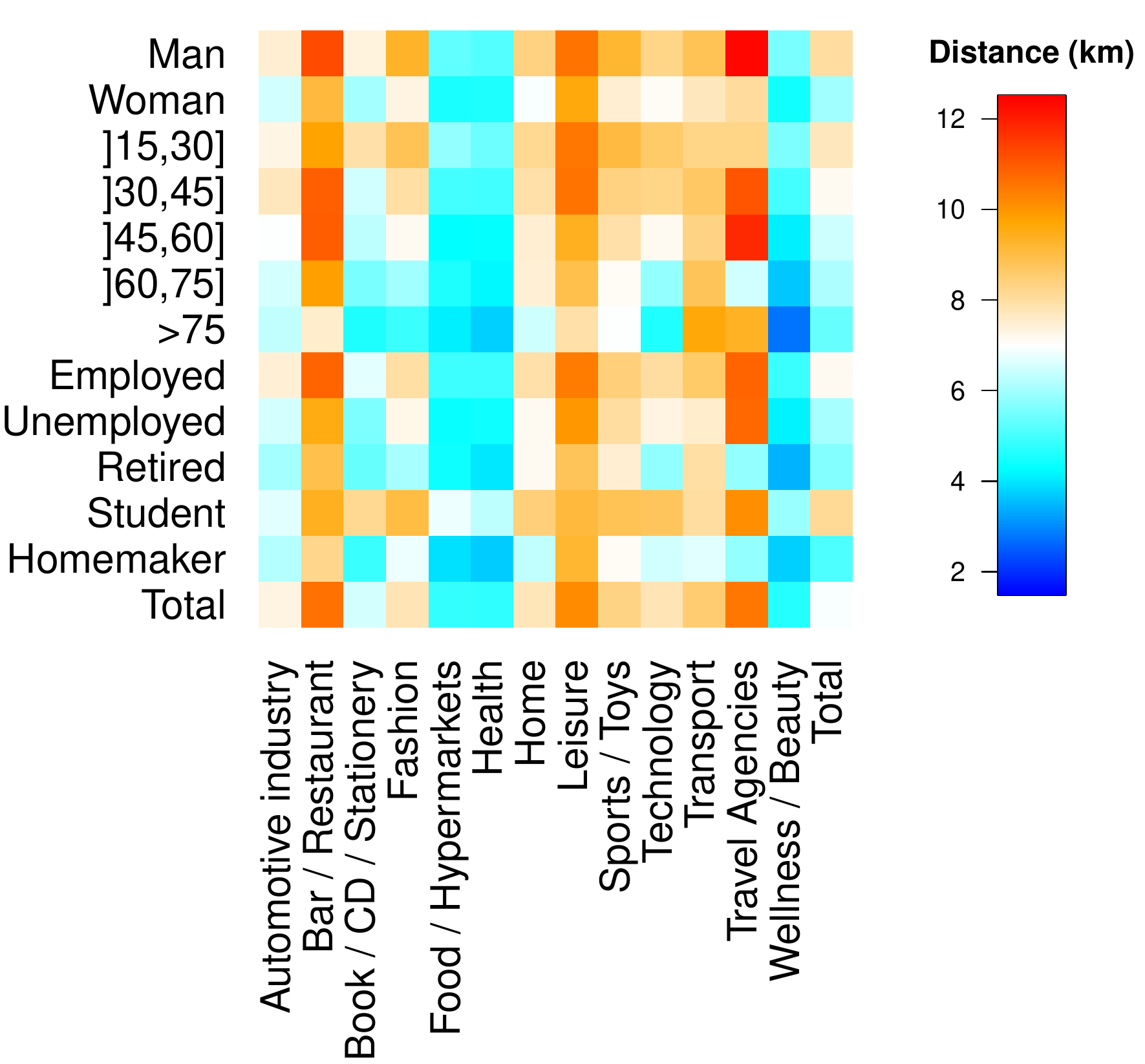}
\caption{\textbf{Average distance between individuals' place of residence and business according to individual's demographics and business' category.} Distances are expressed in kilometer and are computed using the Haversine distance between the latitude and longitude coordinate of the centroid of the customer's postcode of residence and the business' latitude and longitude coordinates. \label{FigS14}}
\end{figure*}

\begin{figure*}
\centering
\includegraphics[width=\linewidth]{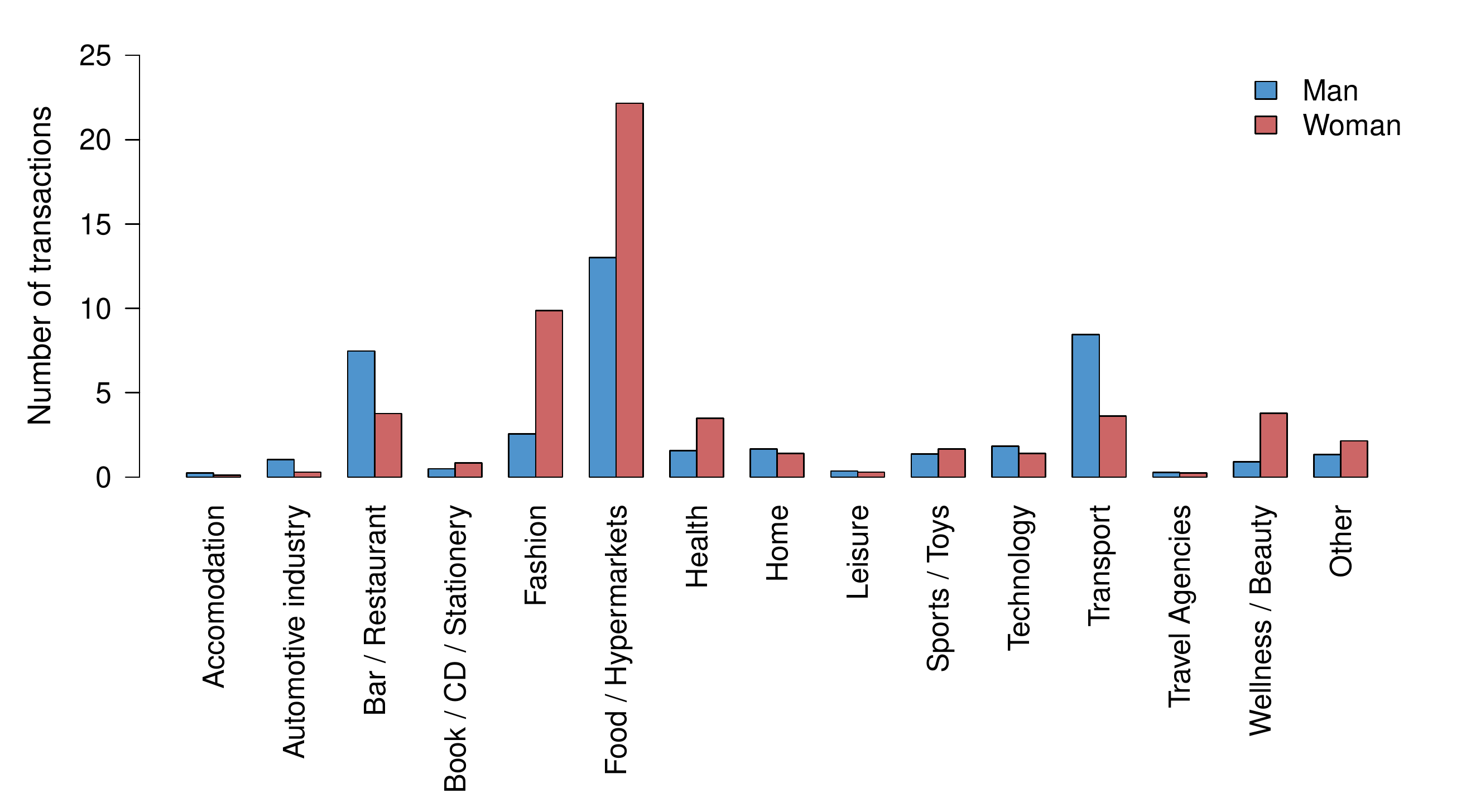}
\caption{\textbf{Average number of transactions according to the gender and the business category.} \label{FigS15}}
\end{figure*}

\end{document}